\documentclass[pra,twocolumn,showpacs,groupedaddress,notitlepage,superscriptaddress,floatfix,nofootinbib]{revtex4-2}

\usepackage{subfigure}
\usepackage{braket}
\usepackage{tikz}
\usepackage{amsmath}
\usepackage{listings}
\usepackage[ruled]{algorithm} 
\usepackage{algpseudocode}
\usepackage{amsfonts}
\usepackage{float,mathdots}

\usepackage{multirow}
\usepackage{array}
\usepackage{amssymb}
\usepackage{accents}
\usepackage{amsthm}

\usepackage{hyperref}
\usepackage{blkarray}
\usepackage{svg}
\usepackage[dvipsnames]{xcolor}
\usepackage{booktabs}
\usepackage{makecell} % for line breaks inside cells

\usepackage{amsmath}

\usepackage{appendix}

\hypersetup{
    colorlinks=true,   % false: boxed links; true: colored links
    linkcolor=cyan,    % color of internal links
    citecolor=magenta, % color of links to bibliography
    filecolor=magenta, % color of file links
    urlcolor=cyan,     % color of external links
    runcolor=cyan
}
\usepackage[capitalise]{cleveref} % Should be loaded after hyperref
\newcommand{\appropto}{\mathrel{\vcenter{
  \offinterlineskip\halign{\hfil$##$\cr
    \propto\cr\noalign{\kern2pt}\sim\cr\noalign{\kern-2pt}}}}}

\begin{document}

\title{Holonomic quantum gates via continuous measurement in bosonic codes: GKP and cat states}

\author{Juan Garcia-Nila}
\affiliation{Department of Electrical \& Computer Engineering, University of Southern California, Los Angeles, California}

\author{Anirudh Lanka}
\affiliation{Department of Electrical \& Computer Engineering, University of Southern California, Los Angeles, California}

\author{Todd A. Brun}
\affiliation{Department of Electrical \& Computer Engineering, University of Southern California, Los Angeles, California}

\begin{abstract}
We apply continuous measurement-based holonomic quantum computation (CMHQC)  \cite{Lanka2025,Lanka2026} to bosonic quantum error-correcting codes and develop explicit protocols for both four-component cat codes and Gottesman–Kitaev–Preskill (GKP) codes. In this framework, a continuously monitored time-dependent codespace undergoes a closed trajectory on the Grassmannian manifold while Zeno confinement suppresses departures from the instantaneous code subspace. For cat codes, we construct a family of squeezed-cat trajectories whose projected Wilczek–Zee connection generates arbitrary logical $Z_L$ rotations, including non-Clifford $T_L$-gates. For GKP codes, we introduce a translated-lattice trajectory that realizes the logical $T_\mathrm{GKP}$ gate through a purely geometric holonomy. We derive the corresponding time-dependent projectors, analytically evaluate the projected connections, and show that the resulting holonomies reproduce the desired logical operations without Hamiltonian control. Furthermore, we analyze the error-correcting capabilities of the instantaneous codespaces by establishing dressed Knill–Laflamme conditions for the relevant bosonic error models and derive analytical estimates for leakage induced by finite-strength continuous measurements. Our results provide a concrete realization of measurement-induced holonomic control in experimentally relevant bosonic platforms and establish a full fault-tolerant logical gate implementation.

\end{abstract}

\maketitle

\section{Introduction}
Fault-tolerant quantum computation requires a universal set of quantum logical gates that preserve compatibility with quantum error correction. In particular, realizing non-Clifford logical gates remains one of the central challenges in developing scalable quantum processors. Conventional approaches typically rely on Hamiltonian engineering, gate teleportation, or magic-state injection, often introducing significant experimental overhead or very stringent control requirements.

An alternative paradigm is provided by holonomic quantum computation, where logical operations arise from geometric phases accumulated during the cyclic evolution of a degenerate subspace rather than from dynamical phases alone \cite{Wilczek_Zee,zanardi_1999,Mommers_2022}. Because geometric transformations depend primarily on the global properties of the evolution path, rather than the exact timing or interaction strength, holonomic protocols have long been regarded as a promising route toward robust quantum control.

Recently, continuous measurement-based holonomic quantum computation (CMHQC) was introduced as a framework in which logical gates emerge from the continuous monitoring of a time-dependent code subspace \cite{Lanka2025,Lanka2026}. In this approach, a family of projectors traces a closed path on the Grassmannian manifold of encoded subspaces, while repeated measurements enforce Zeno confinement to the instantaneous code space. The resulting logical operation is determined by the associated Wilczek--Zee holonomy rather than by explicit Hamiltonian driving.

Despite its conceptual appeal, the applicability of CMHQC to experimentally relevant quantum error-correcting architectures remains largely unexplored. In particular, the original formulation is naturally expressed in terms of stabilizer measurements, and its implementation in conventional qubit codes often requires the measurement of high-weight, many-body Pauli operators (or linear combinations of such operators).

Bosonic quantum error-correcting codes provide a natural setting in which to address this challenge. By encoding quantum information in harmonic oscillators, bosonic codes exploit large Hilbert spaces to achieve hardware-efficient error correction \cite{GKP2001,Mirrahimi_2014}. Among the most highly developed examples are cat codes \cite{Leghtas_2015,Touzard_2018,Grimm_2020} and Gottesman--Kitaev--Preskill (GKP) codes \cite{GKP2001}. The latter convert small displacement errors into correctable syndromes and have become a promising architecture for fault-tolerant bosonic quantum computation \cite{Glancy_2006,Fukui2018}. Recent experimental demonstrations of GKP state preparation and error correction further highlight their practical relevance \cite{Fluhmann2019,deNeeve2025}. Importantly, the stabilizers of bosonic codes correspond to experimentally accessible oscillator observables, such as parity operators and phase-space displacements, making them particularly attractive platforms for implementing CMHQC.

At the same time, geometric and holonomic control of bosonic systems has become an active area of research \cite{Albert_2016,Krastanov_2015,Kang_2022,Zhang_2024,Puri_2020}. Existing bosonic holonomic protocols typically rely on Hamiltonian control of the oscillator manifold through adiabatic deformations or engineered nonlinearities. In contrast, CMHQC generates geometric operations by continuous measurements and Zeno confinement of a time-dependent code subspace.

In this work, we extend CMHQC to bosonic quantum error-correcting codes. We construct explicit holonomic paths for both the four-component cat code and the GKP code, and show how continuous monitoring of the instantaneous codespace generates nontrivial logical operations. For cat codes, we introduce a squeezed-cat trajectory whose projected Wilczek--Zee connection produces arbitrary logical $Z_L$-rotations. For GKP codes, we construct a translated lattice trajectory whose associated holonomy realizes the logical $T_{\mathrm{GKP}}$ gate. Unlike previous bosonic holonomic proposals, our construction generates these operations through continuous measurement rather than direct logical Hamiltonian control.

Beyond the construction of the holonomic paths themselves, we analyze the error-correcting properties of the instantaneous code spaces and derive analytical estimates for leakage induced by finite-strength measurements. These results give a concrete realization of CMHQC in experimentally relevant bosonic platforms and establish a direct connection between continuous-measurement holonomies and bosonic quantum error correction.

The remainder of this paper is organized as follows. In \cref{sec:cat_code} we introduce the cat-code construction and derive the corresponding holonomic logical $Z_L$ rotations. In \cref{sec:GKP} we present the GKP implementation and show how the protocol realizes a logical $T_{\mathrm{GKP}}$ gate. Appendices \ref{app:CMBHQC}--\ref{app:WZ_GKP} contain the geometric framework, derivations of the projected Wilczek--Zee connections, error-correction conditions, and leakage probability analysis used throughout the manuscript.

\section{Cat Code}\label{sec:cat_code}

\begin{figure*}[t]
    \centering
\includegraphics[width=\linewidth]{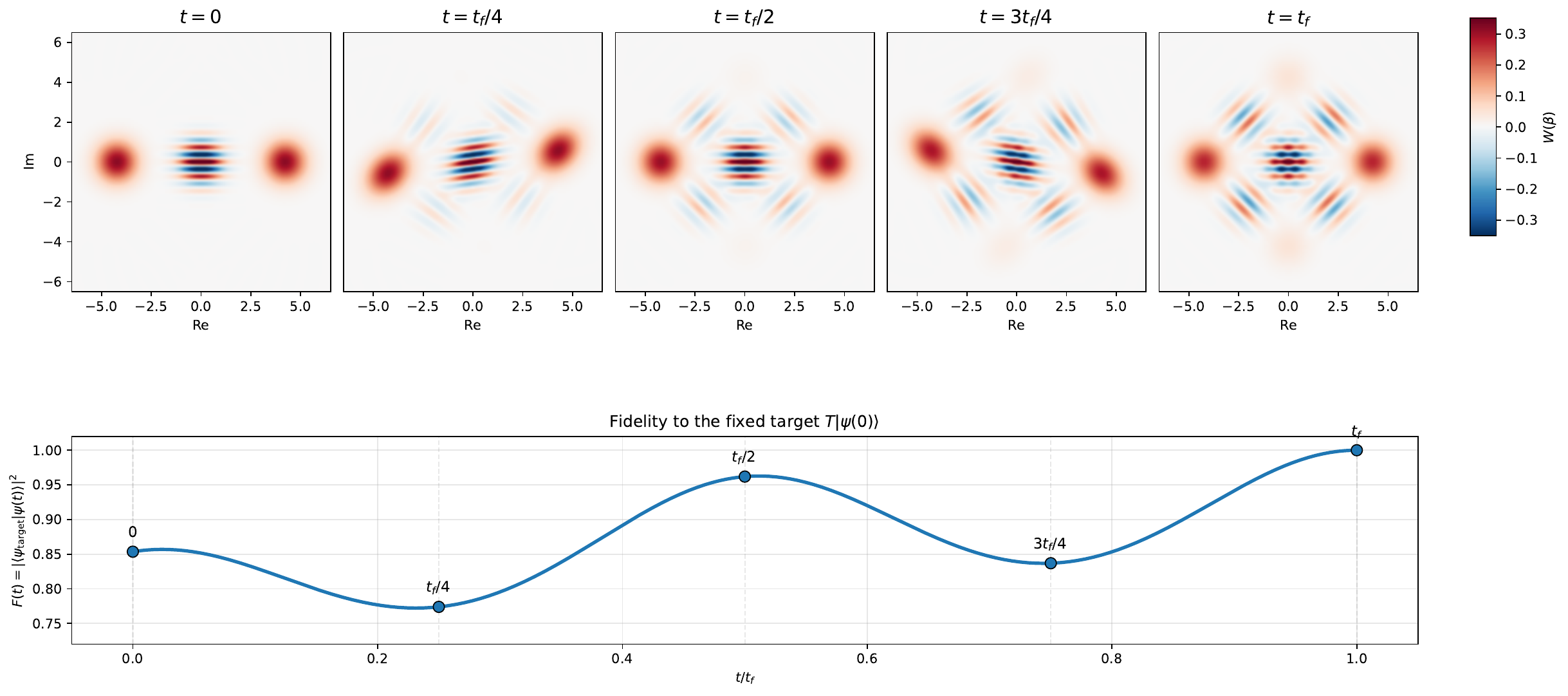}
\caption{Implementation of the logical $T$ gate in the four-component cat code. \textbf{Top:} Wigner-function snapshots of the encoded state $|\psi(t)\rangle=V(t)|\psi(0)\rangle$ at five different times showing its continuous evolution under the holonomic control path. \textbf{Bottom: } Fidelity with respect to the target state $T|\psi(0)\rangle$, where the  $\ket{\psi(0)}=\ket{+}$ is the $+1$ eigenstate with the $X_L$ operator, as a function of the normalized time $t/t_f$. Markers indicate the times corresponding to the Wigner-function snapshots. The protocol reaches unit fidelity at $t=t_f$, demonstrating the implementation of the target logical gate.}
\label{fig:cat_holonomic_gate_evolution}
\end{figure*}

\subsection{Code description}

A natural bosonic encoding is the Schr\"{o}dinger cat code, built from superpositions of coherent states \cite{Mirrahimi_2014,Leghtas_2015}. A coherent state, denoted as $\ket{\alpha}$, is an eigenstate of the annihilation operator $a$:
\begin{equation}\label{eq:coherent}
a\ket{\alpha}=\alpha\ket{\alpha}    
\end{equation}
where $\alpha \in \mathbb{C}$ is the complex eigenvalue of the coherent state $\ket{\alpha}$.

We consider the 4-component cat subspace, spanned by the four cat states
\begin{equation}\label{eq:cat_4}
\ket{C_{\alpha}^k}\equiv \mathcal{N}_k\sum_{m=0}^3 i^{-km}\ket{i^m \alpha}, \qquad \text{for }k=0,1,2,3;
\end{equation}
which form an orthonormal basis with normalization factor
\begin{equation}
\mathcal{N}_k=\frac{1}{4}e^{|\alpha|^2/2}\left(\sum_{n=0}^{\infty}\frac{|\alpha|^{2(4n+k)}}{(4n+k)!}\right)^{-1/2}=\frac{1}{2}+\mathcal{O}(e^{-|\alpha|^2}).
\end{equation}
We can see immediately that these states are degenerate eigenstates of the $a^4$ operator with eigenvalue $\alpha^4$, so the 4-component code manifold is
\begin{equation}\label{eq:ker_a4}
\mathcal{C}_{\alpha}\equiv \ker(a^4-\alpha^4 I).
\end{equation}
While $a^4$ does not distinguish the four states, using the number operator 
\begin{equation}\label{eq:N_op}
    N\equiv a^{\dagger}a ,
\end{equation}
we define the quarter-rotation operator:
\begin{equation}\label{eq:R4_op}
    R_{\pi/2}\equiv e^{i\frac{\pi}{2}N} ,
\end{equation}
which lets us distinguish between the cat states, since
\begin{equation}
R_{\pi/2}\ket{C_{\alpha}^k}=i^k \ket{C_{\alpha}^k}.
\end{equation}

The usual qubit encoding is the \textbf{even-parity} subspace
\begin{subequations}\label{eq:cat_01}
\begin{align}
\label{eq:cat_0}\ket{0_L}&\equiv \ket{C_{\alpha}^0},\\ \label{eq:cat_1}\ket{1_L}&\equiv\ket{C_{\alpha}^2}.
\end{align}
\end{subequations}
Thus, we can define $R_{\pi/2}$ to be the $Z_L$ logical operator, since it acts on the code space just as the encoded Pauli does. If we apply it to the $\ket{0_L}$ and $\ket{1_L}$ states we get
\begin{subequations}\label{eq:ZL_cat_op}
\begin{align}
    Z_L&=R_{\pi/2},\\
    Z_L\ket{0_L}=\ket{0_L}\quad & \qquad Z_L\ket{1_L}=-\ket{1_L}.
\end{align}\end{subequations}

Another important operator is the {\it parity}, which is the half rotation
\begin{equation}\label{eq:parity_op}
    \Pi\equiv e^{i\pi N}=Z_L^2,
\end{equation}
which leaves the code states invariant. We can check that the operator is both hermitian and unitary: $\Pi^{\dagger}=\Pi$ and $\Pi^2=I$.

The annihilation operator \cref{eq:coherent} takes a code state out of the logical subspace. Applied to the cat states in \cref{eq:cat_4}, it yields
\begin{equation}\label{eq:a_over_cat}
a\ket{C_{\alpha}^k} = \alpha \frac{\mathcal{N}_k}{\mathcal{N}_{k-1}} \ket{C_{\alpha}^{k-1}} = \alpha\ket{C_{\alpha}^{k-1}} + \mathcal{O}(e^{-|\alpha|^2}).
\end{equation}
We see that $a$ takes the even-parity sector into the \textbf{odd-parity} sector, and vice versa. Therefore, the parity operator acts as the stabilizer of the even-parity code space:
\begin{subequations}
\begin{align}    
\Pi\ket{0_L}=+\ket{0_L}\quad & \qquad \Pi\ket{1_L}=+\ket{1_L},\\
\Pi\ket{C_{\alpha}^1}=-\ket{C_{\alpha}^1}\quad & \qquad \Pi\ket{C_{\alpha}^3}=-\ket{C_{\alpha}^3}.
\end{align}
\end{subequations}
Since a change of parity will detect if the operator $a$ was applied, we see that photon loss, the physical process associated with $a$, is a correctable error for this code.

The projector onto the 4-cat subspace is 
\begin{equation}
\mathbb{P}_{\text{4cat}}=\sum_{k=0}^3\ket{C_{\alpha}^k}\bra{C_{\alpha}^k},
\end{equation}
and the projector onto the codespace is
\begin{equation}\label{eq:cat_proj}
\mathbb{P}_{0} = \ket{C_{\alpha}^0}\bra{C_{\alpha}^0} + \ket{C_{\alpha}^2}\bra{C_{\alpha}^2} = \frac{1+\Pi}{2} \mathbb{P}_{\text{4cat}}.
\end{equation}

To engineer the logical $X_L$ that transforms $\ket{C_{\alpha}^2}$ into $\ket{C_{\alpha}^0}$ and vice versa, we use the two-photon squeezing generator $H \propto a^2+a^{\dagger\,2}$, since 
\begin{equation}
a^2\ket{C_{\alpha}^k} = \alpha^2 \frac{\mathcal{N}_k}{\mathcal{N}_{k-2}} \ket{C_{\alpha}^{k-2}} = \alpha^2\ket{C_{\alpha}^{k-2}} + \mathcal{O}(e^{-|\alpha|^2}).
\end{equation}
For large $|\alpha|^2\gg 1$, we get
\begin{equation}
a^2\ket{0_L}\approx\alpha^2\ket{1_L},\quad \qquad a^2\ket{1_L}\approx\alpha^2 \ket{0_L}.
\end{equation}
Hence, over the codespace,
\begin{equation}\label{eq:XL_cat}
\mathbb{P}_0(a^2+a^{\dagger\, 2})\mathbb{P}_0\propto X_L .
\end{equation}

There are several alternative methods to generate these Clifford logical operators experimentally using controlled displacements, Selective Number-dependent Arbitrary Phase (SNAP) gates \cite{Heeres2015}, or driven Kerr evolution. However, a universal gate set requires a non-Clifford gate such as
\begin{equation}
T_L = e^{-i\frac{\pi}{8}Z_L} ,
\end{equation}
We propose using the measurement-based holonomic protocol to generate an arbitrary rotation over $Z_L$.

\subsection{Holonomic evolution for cat code}

Before discussing  the details of the holonomic protocol, we define the two-photon squeezing operator, which is the Gaussian unitary
\begin{equation}\label{eq:squeezing_op}
S(\xi) \equiv \exp\left[\frac{1}{2}(\xi^* a^2-\xi a^{\dagger \, 2})\right] ,
\end{equation}
with $\xi\equiv re^{i\vartheta }$, where $r$ and $\vartheta$ are the squeezing strength and phase angle, respectively. This is a standard bosonic operation and the usual implementation is via parametric driving, with the parameters $r$ and $\vartheta$ controlled with the pump power, pump phase and pulse duration.

Now, let us define $V(t)$, which describes the holonomic path that rotates the original codespace $\mathcal{C}_\alpha$ in \cref{eq:ker_a4} to the instantaneous vector space $S(\xi(t))\mathcal{C}_\alpha$. We define $V(t)$ in terms of the quarter rotation operator in \cref{eq:ZL_cat_op} and the squeezing operator in \cref{eq:squeezing_op} as follows:
\begin{equation}\label{eq:rotating_V}
V(t) = e^{i\phi(t)Z_L}S(\xi(t)) ,
\end{equation}
where $\phi(t)$ and $\xi(t)$ are functions that we control. At time $t=0$ the unitary $V$ is simply the identity, but at the final time $t_f$ it will produce the desired logical gate parametrized by the phase $\theta$:
\begin{subequations}\label{eq:V_ini_fin}
\begin{align}
V(0) &= I , \\
V(t_f) &= \exp(i\theta Z_L).
\end{align}
\end{subequations}
Without loss of generality, we propose to use the function
\begin{equation}\label{eq:phi_of_t}
\phi(t) = \frac{\theta}{t_f}t ,
\end{equation}
which satisfies $\phi(0)=0$ initially and $\phi(t_f)=\theta$ at the final time $t_f$.

In addition, we require that the squeezing vanishes at the initial and final times, so we choose 
\begin{equation}\label{eq:xi_cond}
\xi(0)=\xi(t_f)=0. 
\end{equation}

In Appendix~\ref{app:h_t_cat} we show that the following function produces the desired logical operation:
\begin{subequations}\label{eq:xi}
\begin{align}
\xi(t) &\equiv r(t)\exp[i\vartheta(t)] \\
r(t) &= \frac{j_{0,1}}{2|\alpha|^2} \sin\left(\frac{2\pi t}{t_f}\right) \\
\vartheta &= 2\arg(\alpha) - \frac{\pi}{2},
\end{align}
\end{subequations}
where $j_{0,1}\approx 2.4048$ is the first zero root of $J_0$, the Bessel function of the first kind.

We transform the initial cat-code projector given in \cref{eq:cat_proj} into the projector for the instantaneous codespace by the unitary in \cref{eq:rotating_V}:
\begin{equation}\label{eq:projector_t}
\mathbb{P}(t)=V(t)\mathbb{P}_{0} V^{\dagger}(t).
\end{equation}
The associated stabilizer is
\begin{equation}\label{eq:Pi_t_cat}
\Pi(t) \equiv V(t)\Pi V^{\dagger}(t) = \Pi.
\end{equation}
This time-dependent projector gives a way to maintain the state in the instantaneous codespace at all times while following the holonomic path, which produces the desired gate at the final time $t_f$ unless a measurement error occurs.

In the next section we discuss the probability of leakage due to a measurement leaving the state in the orthogonal complement of the code space with rotated projector $I-\mathbb{P}(t)$. In the adiabatic limit of long $t_f$ we expect the probability of staying in the code to approach 1.

\cref{fig:cat_holonomic_gate_evolution} illustrates the implementation of the logical gate $T_L$ in the four-component cat code. The simulation starts from the logical state $|+\rangle_L = (|0_L\rangle + |1_L\rangle)/\sqrt{2}$, encoded in the harmonic oscillator Hilbert space. The state evolves according to the time-dependent unitary path $V(t)$ in \cref{eq:rotating_V}, using $\theta=\pi/8$, $\alpha=3$, $t_f=5$, and a Fock-space truncation of $N_{\max}=40$. During the evolution, the coherent-state components undergo simultaneous squeezing and rotation in phase space, producing a continuous deformation of both the Gaussian peaks and the interference fringes, as seen in the Wigner-function snapshots. In the lower panel we show the fidelity, which begins at $F(0) = \bra{+}T\ket{+} = \cos^2(\pi/8)$ and reaches unit fidelity at $t=t_f$.

\begin{figure}
\centering
\includegraphics[width=\linewidth]{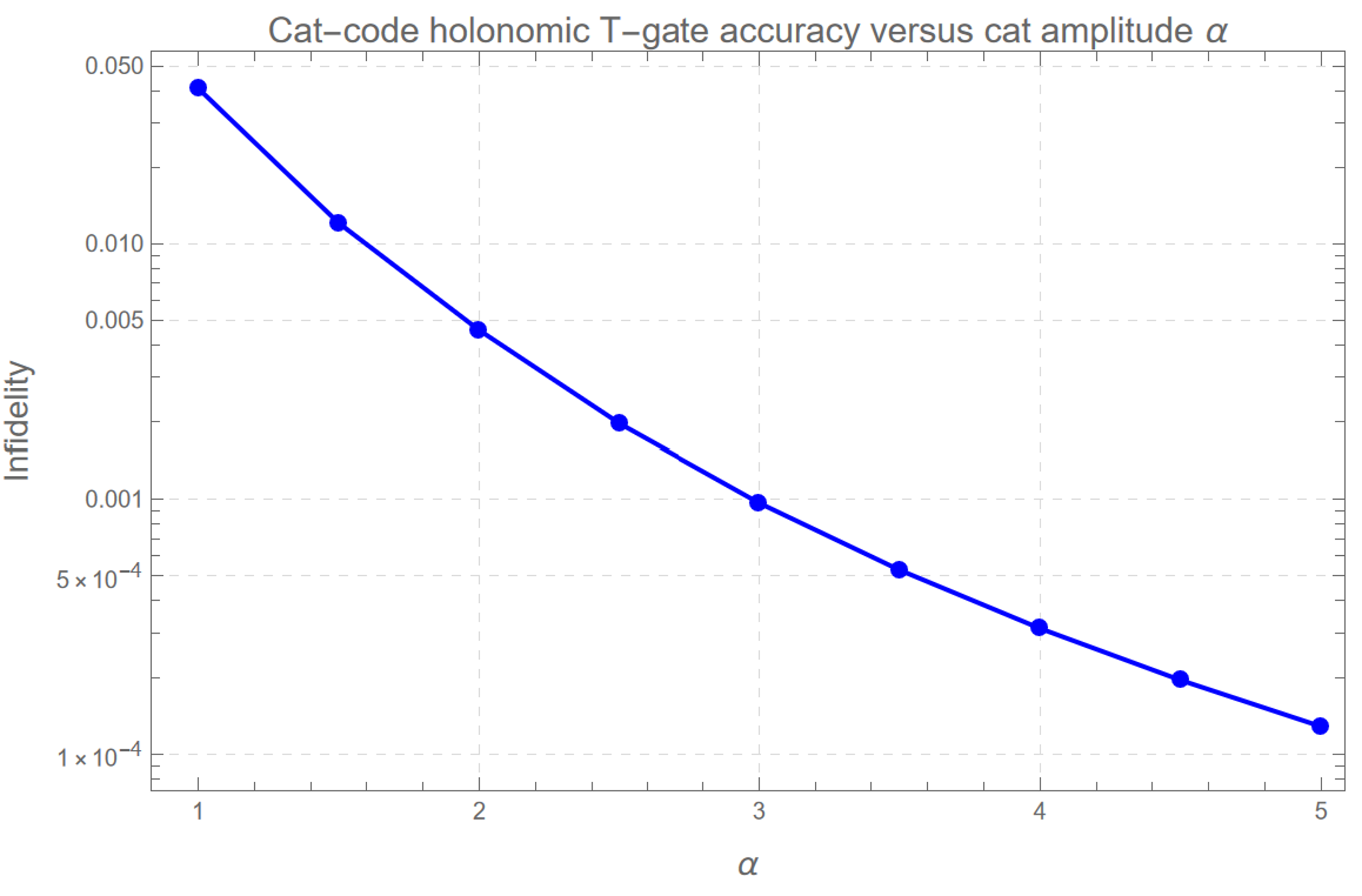}
\caption{Process holonomy infidelity $1 - {\left| \mathrm{Tr} \left (T_L^\dagger  U_ {\mathrm {hol}} \right) \right|^2}/{4}$ of  the  holonomic  logical gate $T_L = e^{i\pi  Z_L/8}$, as  a  function  of  the  cat-state  amplitude  $|\alpha|$. The  logical  holonomy $U_{\mathrm {hol}}$  is  obtained  numerically. The  gate  performance  is  quantified  by  the  process  fidelity. For  $|\alpha|\gtrsim  3$, the  process  infidelity  falls  below  $10^{-3}$, demonstrating a high-fidelity implementation of  the logical non-Clifford gate in the large-$|\alpha|$ limit.}
\label{fig:cat_infidelity}
\end{figure}

\subsection{Knill-Laflamme conditions}

In this subsection, we analyze whether the cat code defined in \cref{eq:ker_a4} and the rotated code defined by the projector in \cref{eq:projector_t} satisfy the Knill–Laflamme (KL) conditions for error correction. 

For the correctable error set $\mathcal{E} = \{I,a\}$, we show in Appendix~\ref{app:KL_gkp} that the following conditions hold:
\begin{subequations}\label{eq:KL_cat}
\begin{align}
\mathbb{P}_0 I\mathbb{P}_0 &= \mathbb{P}_0, \\ \label{eq:a_corr_cat}
\mathbb{P}_0 a\mathbb{P}_0 &= 0, \\ \label{eq:a_dag_corr_cat}
\mathbb{P}_0 a^{\dagger}\mathbb{P}_0 &= 0, \\ \label{eq:N_cat}
\mathbb{P}_0 a^{\dagger}a\mathbb{P}_0 &= N\mathbb{P}_0, \\ \label{eq:Nplus1_cat}
\mathbb{P}_0 aa^{\dagger}\mathbb{P}_0 &= (N+I)\mathbb{P}_0,
\end{align}
\end{subequations}
for $\mathbb{P}_0$ in \cref{eq:cat_proj}. The condition for the trivial error $I$ is immediate, since $\mathbb{P}_0^2=\mathbb{P}_0$, and \cref{eq:Nplus1_cat} follows from \cref{eq:N_cat}, recognizing that $N=a^{\dagger}a=aa^{\dagger}-I$. The other properties are shown in Appendix~\ref{app:KL_cat_code}.

So far, the conditions \cref{eq:KL_cat} implies that only the error detection conditions hold. However, in the large-amplitude limit $|\alpha|^2\gg 1$, the error conditions hold to a very good approximation:
\begin{subequations}\label{eq:KL_Nop_cat}
\begin{align}
\mathbb{P}_0 a^{\dagger} a \mathbb{P}_0 &= |\alpha|^2 \mathbb{P}_0 + \mathcal{O}(e^{-|\alpha|^2}), \\
\mathbb{P}_0 a a^{\dagger} \mathbb{P}_0 &= (|\alpha|^2 + 1) \mathbb{P}_0 + \mathcal{O}(e^{-|\alpha|^2}).
\end{align}
\end{subequations}

Now we need to show that the time-dependent projector $\mathbb{P}(t)$ in \cref{eq:projector_t}
obeys the KL conditions in \cref{eq:KL_cat} in substitution from $\mathbb{P}_0$, equivalent to
\begin{subequations}\label{eq:KL_t_cat}
\begin{align}\label{eq:KL_Pt_a}
\mathbb{P}(t) a \mathbb{P}(t) &= 0, \\ \label{eq:KL_adag_elxokas}
\mathbb{P}(t) a^{\dagger} \mathbb{P}(t) &= 0, \\ \label{eq:KL_N_Eleazar}
\mathbb{P}(t) N\mathbb{P}(t) &= \left[\cosh(2r)|\alpha|^2 + \sinh^2(r)\right] \mathbb{P}(t) + \mathcal{O}(e^{-|\alpha|^2}).
\end{align}
\end{subequations}
These properties are shown in \cref{app:KL_cat_code} by defining the time dependent annihilator and number operators due to the unitary \cref{eq:rotating_V}
\begin{subequations}
\begin{align}\label{eq:a_of_t}
a(t) &\equiv V^{\dagger}(t) a V(t), \\ \label{eq:a_dagger_of_t}
a^{\dagger}(t) &= V^{\dagger}(t) a^{\dagger} V(t), \\ \label{eq:N_of_t}
N(t) &\equiv V^{\dagger}(t) a^{\dagger} a V(t).
\end{align}
\end{subequations}

\begin{figure}
\centering
\includegraphics[width=\linewidth]{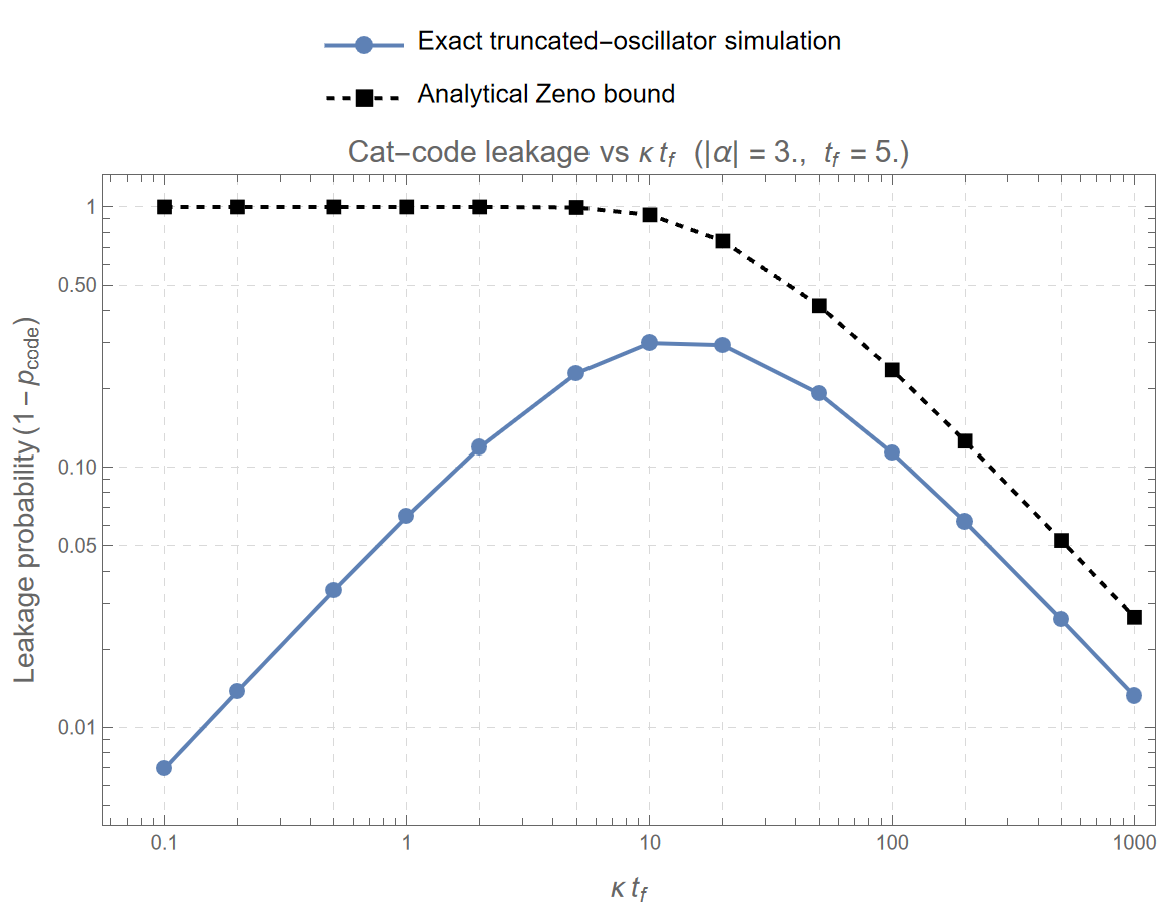}
\caption{
Leakage probability $1-p_{\mathrm{code}}$ of the cat-code holonomic $T_L$ gate as a function of the dimensionless measurement strength $\kappa t_f$ for $|\alpha|=3$ and $t_f=5$. The blue curve shows the exact truncated-oscillator simulation obtained from the Lindblad master equation in the rotating frame, while the black dashed curve denotes the analytical operator-norm upper bound. The plotted quantity is the leakage infidelity.} %For sufficiently large $\kappa t_f$, the leakage probability decreases, demonstrating the effectiveness of continuous measurement in confining the dynamics to the cat-code manifold.}
\label{fig:leakage_cat}
\end{figure}

\subsection{Leakage probability due to measurement}

Finally, we discuss the probability of leakage out of the instantaneous codespace at the end of the evolution, due to the constant measurement of the instantaneous projector in \cref{eq:projector_t}.

The dynamics of the state vector $\ket{\psi(t)}$ of the system has a stochastic description, due to the randomness of quantum measurements. We can find an average description of the state at a finite time $t_f$ by using the ensemble average over all measurement outcomes, which gives a Lindblad equation for the density matrix $\rho=\mathbb{E}[\ket{\psi}\bra{\psi}]$:
\begin{equation}\label{eq:Lindblad_cat_code}
\frac{d\rho(t)}{dt} = \kappa \mathcal{D}[\mathbb{P}(t)]\rho(t).
\end{equation}
A useful trick to solve \cref{eq:Lindblad_cat_code} consists in switching to the rotating frame, defining
\begin{equation}
\tilde{\rho}(t)\equiv V^{\dagger}(t)\rho(t) V(t),
\end{equation}
so the equation becomes
\begin{equation}\label{eq:Lindblad_RF_cat_code}
\frac{d \tilde{\rho}(t)}{dt}=-[V^\dagger(t)\dot{V}(t),\tilde{\rho}(t)]+\kappa \mathcal{D}[\mathbb{P}_0]\tilde{\rho}(t) .
\end{equation}

We can interpret the Wilczek-Zee connection $-iV^{\dagger}\dot V$ as an effective Hamiltonian. Hence, we can compute the probability of staying in the codespace after the whole evolution by measuring the rotated projector in \cref{eq:projector_t} at a rate $\kappa$. We decompose the Hilbert space into the codespace $\mathcal{C}$, defined with the basis in \cref{eq:cat_01}, and the leakage subspace $\mathcal{C}^{\perp}$:
\begin{equation}\label{eq:H_C_Cperp}
\mathcal{H}=\mathcal{C}\oplus\mathcal{C}^{\perp} .
\end{equation}
We define the leakage block as 
\begin{equation}\label{eq:B_ortho}
B(t) \equiv \mathbb{Q}_0 V^{\dagger}(t) \dot{V}(t) \mathbb{P}_0
\end{equation}
where $\mathbb{Q}_0$ is the projector onto the orthogonal complement of the codespace, defined as
\begin{equation}\label{eq:Q0_ortho}
    \mathbb{Q}_0=I-\mathbb{P}_0.
\end{equation}
In Appendix~\ref{app:Leak_prob_cat} we show that the leakage probability is bounded 
\begin{equation}
\label{eq:leak_prob_cat}
    1-p_{\mathrm{code}}(t_f)\leq 1-\exp\left(-\frac{8}{\kappa}\int_{0}^{t_f}||B(t )||_2^2 dt\right).
\end{equation}

We can easily check that this probability will be close to zero in the \textit{adiabatic limit} of a long protocol time $t_f$ with a strong measurement rate $\kappa \gg 1$.

\section{GKP code}\label{sec:GKP}

\begin{figure*}[t]
\centering
\includegraphics[width=\linewidth]{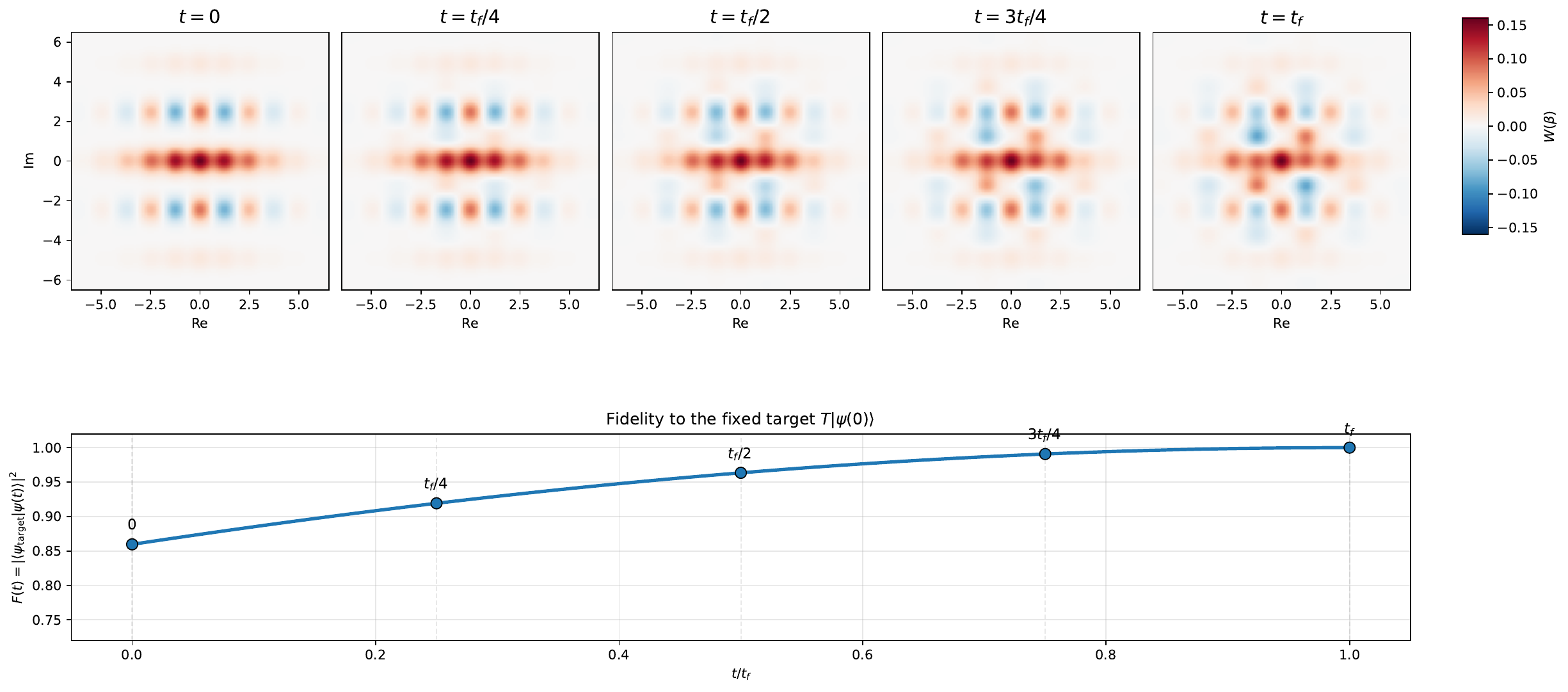}
\caption{Measurement-based holonomic implementation of the $T_{\mathrm{GKP}}$ gate on the GKP code. \textbf{Top:} Wigner function snapshots of the encoded state at five different times. \textbf{Bottom:} Fidelity with respect to the target state $T\ket{\psi(0)}$ as a function of the rescaled time $t/t_f$, where the initial state $\ket{\psi(0)}=\ket{+}$ is the $+1$ eigenstate of the $X_L$ operator. Markers indicate the times corresponding to the Wigner-function snapshots above.}    
\label{fig:gkp_holonomic_gate_evolution_sf_soft}
\end{figure*}

\subsection{Code description}

The GKP code is a quantum error-correcting code that encodes a logical qubit in a single bosonic mode. It is named after the seminal paper published in 2001 by Daniel Gottesman, Alexei Kitaev and John Preskill \cite{GKP2001}.

It is useful to define the displacement operator as a function of an arbitrary complex number $\alpha$:
\begin{equation}\label{eq:displacement}
D(\alpha)\equiv\exp(\alpha a^{\dagger}-\alpha^* a).
\end{equation}
The important properties of this operator are given in Appendix~\ref{app:displace}. In particular, the name comes from the property given in \cref{eq:displace_prop}. In addition, it is useful to define the position and momentum quadratures, respectively:
\begin{subequations}\label{eq:quadratures}
\begin{align}\label{eq:Q_quad}
Q &\equiv \frac{i}{\sqrt{\pi}}(\beta^*a-\beta a^{\dagger}),\\
P &\equiv -\frac{i}{\sqrt{\pi}}(\alpha^*a-\alpha a^{\dagger}),\\
& [Q,P]=i.
\end{align}
\end{subequations}
The logical Pauli operators on the code are defined in terms of displacement operators:
\begin{subequations}\label{eq:logicals_GKP}
\begin{align}\label{eq:X_GKP}
X_L&\equiv D(\alpha)=e^{-i\sqrt{\pi}P},\\\label{eq:Z_GKP}
    Z_L&\equiv D(\beta)=e^{i\sqrt{\pi}Q},
\end{align}
\end{subequations}
where, to satisfy the anticommutation property of the operators, according to \cref{eq:Comm_Disp} we must select $\alpha$ and $\beta$ such that
\begin{equation}
    \beta \alpha^*-\beta^* \alpha = i\pi.
\end{equation}
Furthermore, with the property in \cref{eq:YL_Disp}, we can define the logical operator $Y_L$ as
\begin{equation}
    Y_L = i X_L Z_L = D(\alpha+\beta)=e^{i\sqrt{\pi}(Q-P)}.
\end{equation}

To complete the definition of the logical codespace, we define the stabilizer generators as
\begin{subequations}\label{eq:S_GKP}
\begin{align}\label{eq:SX_GKP}
    S_X\equiv X_L^2=D(2\alpha)=e^{-i2\sqrt{\pi}P},\\\label{eq:SZ_GKP}
    S_Z\equiv Z_L^2=D(2\beta)=e^{i2\sqrt{\pi}Q}.
\end{align}
\end{subequations}
They are chosen so that the squares of the logical Pauli operators leave the codespace invariant. The stabilizer group of the code is 
\begin{equation}
\{S_X^k,S_Z^l\} \quad k,l\in \mathbb{Z}.
\end{equation}
The basis of the logical codespace is supported on even and odd integer multiples of $\sqrt{\pi}$ in the position basis,
\begin{subequations}\label{eq:0_1_GKP}
\begin{align}
    \ket{{0}_L}&\equiv\sum_{j=-\infty}^{\infty}\ket{2j\sqrt{\pi}}_Q,\\
    \ket{{1}_L}&\equiv\sum_{j=-\infty}^{\infty}\ket{(2j+1)\sqrt{\pi}}_Q.
\end{align}
\end{subequations}
These states are $\pm1$ eigenstates of $Z_L$ and $+1$ eigenstates of $S_X$ and $S_Z$.

Logical Clifford gates in the GKP code are given by Gaussian unitary operators, i.e., operators with an exponent that is a quadratic function of $q$ and $p$. For instance, the Hadamard, phase and CNOT gates are given by
\begin{subequations}\label{eq:Clifford_GKP}
\begin{align}
    H_L&=e^{i\frac{\pi}{4}(Q^2+P^2)},\\
    S_L&=e^{\frac{i}{2}Q^2},\\
    CX_L&=e^{-i\,PQ}.
\end{align}
\end{subequations}

% ===== GKP holonomic path =====

While logical Clifford gates use Gaussian unitary operators, non-Clifford gates require non-Gaussian operations. However, non-Clifford gates are necessary for universality. In this paper, we propose a way to create the $\pi/8$-phase gate, also known as the $T$ gate, using the measurement-based holonomic protocol. We are using the original proposal for constructing the gate using a cubic phase gate \cite{GKP2001}, difficult but experimentally achievable methods to create an accurate $T-$gate had been proposed, like noise-biasing techniques \cite{Nguyen2026}.

The original GKP \cite{GKP2001} proposal for the $T$ gate is
\begin{subequations}\label{eq:T_GKP}
\begin{align}
T_{\text{GKP}} = &\exp\Bigg\{i2\pi \left[\frac{1}{4} \left(\frac{Q}{\sqrt{\pi}} \right)^3 \hspace{-0.1cm} + \frac{1}{8} \left(\frac{Q}{\sqrt{\pi}} \right)^2 \hspace{-0.1cm} - \frac{1}{4} \left(\frac{Q}{\sqrt{\pi}} \right) \right] \Bigg\} \\
=& \exp\left[ i \, 2\pi f \left( \frac{Q}{\sqrt{\pi}} \right) \right] ,
\end{align}
\end{subequations}
where $f$ is the cubic polynomial
\begin{align}\label{eq:f_gkp}
f(x) = \frac{1}{4}x^3 + \frac{1}{8}x^2 - \frac{1}{4}x.
\end{align}
This function has the property that if given an even integer $x$ as an input, the output $f(x)$ is also an integer, while if given an odd integer $x$ then $f(x)=n+1/8$ for some integer $n$. Therefore, applying the gate $T_{\text{GKP}}$ to the even and odd states of the position basis gives
\begin{subequations}
\begin{align}
T_{\text{GKP}}\ket{2n\sqrt{\pi}}_Q &= \ket{2n\sqrt{\pi}}_Q \\
T_{\text{GKP}}\ket{(2n+1)\sqrt{\pi}}_Q &= e^{i\pi/4}\ket{(2n+1)\sqrt{\pi}}_Q,
\end{align}
\end{subequations}
which is valid for any integer $n$. So we conclude that this gate applies the $\pi/8$ Z rotation to the codespace, following the definitions of the logical basis in \cref{eq:0_1_GKP}.

\subsection{Holonomic evolution for the GKP code}

Let us define $V(t)$ to be the unitary describing the holonomic path to perform the non-Clifford gate $T_{\text{GKP}}$ in \cref{eq:T_GKP}. $V(t)$ is parametrized by the time $t$, and it rotates the whole codespace $\mathcal{C}$:
\begin{equation}\label{eq:man_of_the_year}
V(t) = e^{i \, \kappa(t)\, f(Q/\sqrt{\pi})} e^{-i\sqrt{\pi}g(t)P} ,
\end{equation}
where $f$ is the polynomial given in \cref{eq:f_gkp}, which can be generalized to be any other $(Q,P)$ function to create the desired gates in \cref{eq:Clifford_GKP}, while $g(t)$ and $\kappa(t)$ are functions that we control to satisfy
\begin{subequations}\label{eq:V_ini_fin}
\begin{align}
V(0) &= I \\
V(t_f) &= e^{2i\pi f(Q/\sqrt{\pi})} = T_{\text{GKP}}.
\end{align}
\end{subequations}
We choose the endpoints of the functions to satisfy the following conditions:
\begin{subequations}\label{eq:cond_gkp}
\begin{align}\label{eq:g_conds}
g(0) &= g(t_f) = 0, \\ \label{eq:kappa_conds}
\kappa(0) = 0, &\qquad \kappa(t_f) = 2\pi.
\end{align}
\end{subequations}

The parallel transport condition \cref{eq:horizontal_cond} will help us choose specific functions $\kappa$ and $g$. In particular, in Appendix~\ref{app:WZ_GKP} we show why we choose the functions 
\begin{subequations}
    \begin{align}\label{eq:kappa_t}
        \kappa(t) &= \frac{2\pi t}{t_f}, \\ \label{eq:g_t}
        g(t) &= A \sin^2 \left(\frac{\pi t}{t_f}\right),
    \end{align}
\end{subequations}
where $A=\mp(8\pm 2\sqrt{7})/9$, and for which any choice of sign for $A$ works.

Moreover, the unitary in \cref{eq:man_of_the_year} drives the evolution by modifying the codespace stabilized by \cref{eq:S_GKP}, to that stabilized by the time-dependent stabilizer generators
\begin{equation}
    S_j(t)\equiv V(t)S_jV^{\dagger}(t).
\end{equation}
For the $S_Z$ in \cref{eq:SZ_GKP} in particular, we have
\begin{align}
S_Z(t) &= e^{-i\sqrt{\pi}g P} e^{i2\sqrt{\pi}Q} e^{i\sqrt{\pi}g P} \nonumber\\
&= e^{i 2\sqrt{\pi} (Q+\sqrt{\pi}g(t))} = e^{i 2\pi g(t)} S_Z,
\end{align}
where in the second equality we used the property $e^{-iaP} Q e^{iaP} = Q + a$. For the stabilizer $S_X$ in \cref{eq:SX_GKP} we have
\begin{align}
S_X(t) &= e^{i\,\kappa(t)\,[f(Q/\sqrt{\pi}) - f(Q/\sqrt{\pi} - 2)]}\,S_X \\
&= \exp\!\Bigg\{i\frac{\kappa(t)}{2} \left[3\left(\frac{Q}{\sqrt{\pi}} \right)^2 - 5 \left(\frac{Q}{\sqrt{\pi}}\right) + 2 \right] \Bigg\} S_X, \nonumber
\end{align}
where we used the property
\[
e^{iF(Q)} e^{-iaP} e^{-iF(Q)} = e^{i[F(Q) - F(Q-a)]} e^{-iaP} .
\]
We can easily check that the polynomial $3x^2-5x+2$ is always an even integer when $x$ is an integer; hence, at times $0$ and at $t_f$ we recover the original stabilizer.

The instantaneous projector onto the GKP codespace becomes
\begin{equation}\label{eq:P_GKP_t}
\mathbb{P}_{\mathrm{GKP}}(t) = V(t) \mathbb{P}_{\mathrm{GKP}} V^{\dagger}(t) = \frac{1}{4} (I+S_X(t)) (I+S_Z(t)).
\end{equation}
\cref{fig:gkp_holonomic_gate_evolution_sf_soft} illustrates the evolution of a finite-energy GKP qubit during the implementation of the logical $T_{\mathrm{GKP}}$ gate. The top panels display the Wigner function of the encoded state at five representative times, showing how the lattice structure characteristic of the GKP code is preserved throughout the protocol while the interference pattern gradually changes as the relative logical phase is accumulated. The bottom panel plots the fidelity with respect to the target state $T_{\mathrm{GKP}}\ket{\psi(0)}$, which increases monotonically from $F(0)=\cos^2(\pi/8)$ at $t=0$ to unity at the final time. Despite the noticeable modification of the interference fringes, the overall lattice structure remains unchanged, illustrating that the logical operation modifies the encoded quantum information while preserving the underlying GKP code space.

\subsection{Knill-Laflamme conditions}

In this section, we present the error detecting and correcting conditions for the code during this procedure. We consider displacement errors:
\begin{equation}\label{eq:error_GKP}
E_{i}\equiv e^{-i u_i P}e^{iv_i Q} ,
\end{equation}
where, for integers $m,n$, the ideal square GKP code satisfies the projected conditions
\begin{align}\label{eq:KL_GKP_0}
&\qquad\qquad\qquad\qquad \mathbb{P}_\mathrm{GKP} E_j^{\dagger} E_i \mathbb{P}_{\mathrm{GKP}} \\
&=\begin{cases}
\mathbb{P}_\mathrm{GKP} & i=j,\\
0 &u_i-u_j\neq m\sqrt{\pi},\, v_i-v_j\neq n\sqrt{\pi}.,\\
e^{i\phi_{ij}}X_L^{m}Z_L^n \mathbb{P}_{\mathrm{GKP}} &u_i-u_j=m\sqrt{\pi},\, v_i-v_j=n\sqrt{\pi}.
\end{cases}\nonumber
\end{align}
Hence, if the two errors are the same, you get the identity on the code space; if their difference is not a GKP lattice vector, the sandwich vanishes in the ideal code; and if their difference is exactly a lattice vector, you get a logical Pauli, an uncorrectable error.

For the \textbf{correctable} small-displacement set inside one Voronoi cell,
\begin{equation}\label{eq:Voronoi}
|u_i-u_j| < \sqrt{\pi},\qquad |v_i-v_j|<\sqrt{\pi} ,
\end{equation}
and with distinct errors, the difference is not on the lattice, so $\mathbb{P}_\mathrm{GKP} E_j^{\dagger} E_i \mathbb{P}_{\mathrm{GKP}} = 0$. This is the Knill-Laflamme statement for error correction. To show that our holonomic protocol is fault tolerant, we need to satisfy similar conditions but considering the dressed error operators instead:
\begin{equation}
\tilde{E}_i(t)\equiv V^{\dagger}(t)E_iV(t).
\end{equation}
In Appendix~\ref{app:KL_gkp} we generalize the Knill-Laflamme conditions \cref{eq:KL_GKP_0} to the case of the instantaneous codespace projector, using \cref{eq:P_GKP_t}:
\begin{equation}
\mathbb{P}_{\mathrm{GKP}}(t) E_j^{\dagger} E_i \mathbb{P}_{\mathrm{GKP}}(t),
\end{equation}
which is equivalent to proving the left-hand side of the KL conditions
\begin{equation}
\mathbb{P}_{\mathrm{GKP}} \tilde{E}_j^{\dagger} \tilde{E}_i \mathbb{P}_{\mathrm{GKP}},
\end{equation}
where $\{\tilde{E}\}$ are the error operators transformed by the instantaneous unitary $V(t)$.

\subsection{Truncated GKP states}

\begin{figure}
\centering
\includegraphics[width=\linewidth]{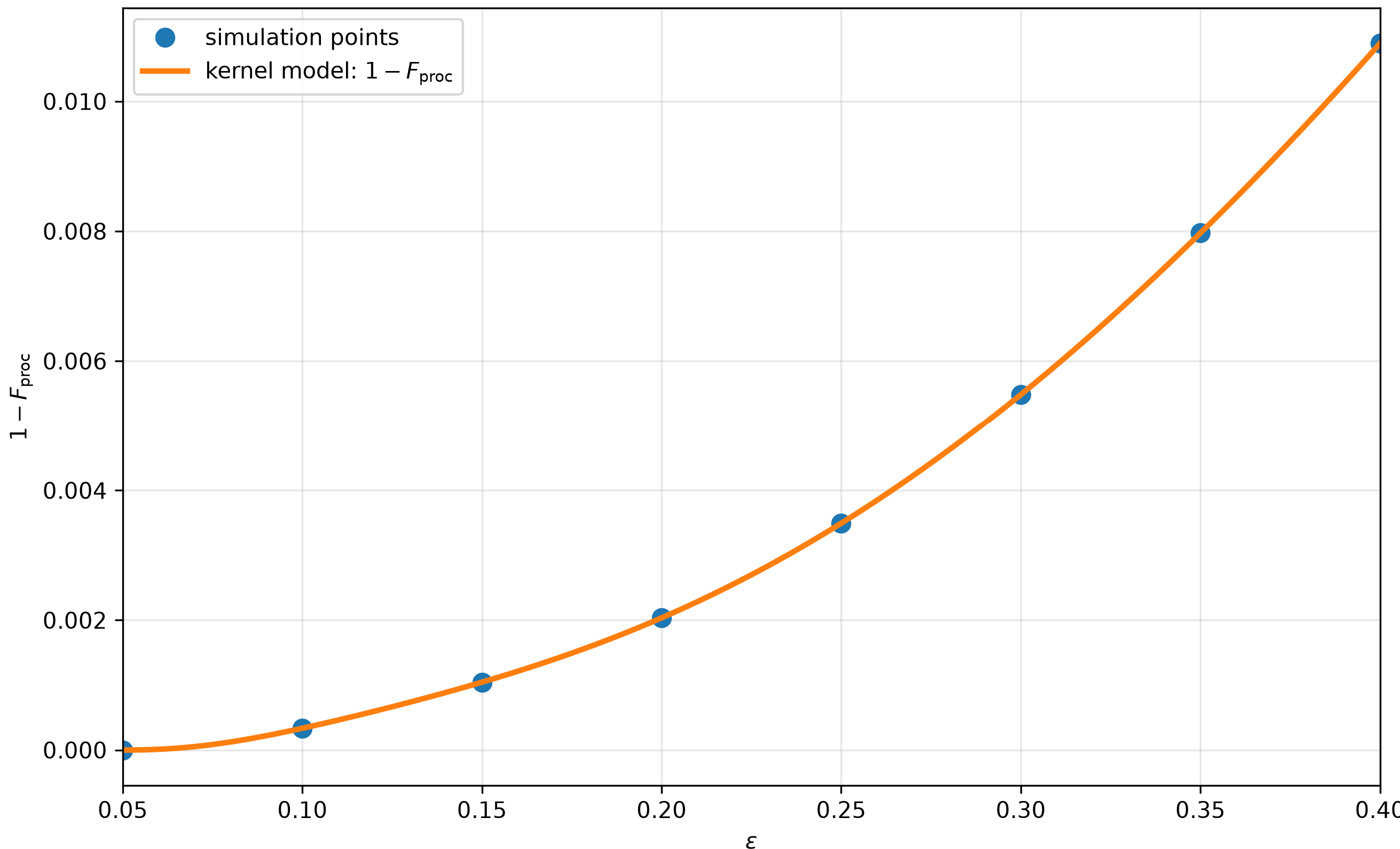}
\caption{Process infidelity $1-F_{\mathrm{proc}}$ of the holonomic GKP $T_{\mathrm{GKP}}$ gate as a function of the finite-energy parameter $\epsilon$. Blue markers show the numerical values using the truncated finite-energy kernel model on Strawberry Fields, while the orange curve is the corresponding smooth interpolation.}
\label{fig:proc_infid_GKP}
\end{figure}

The original idea for defining the GKP code was to use the even and odd harmonic oscillator levels to define the logical states $\ket{0_L}$ and $\ket{1_L}$ in \cref{eq:0_1_GKP}, respectively. However, in practice it is impossible to use the ideal GKP states either in experimental or numerical settings. Hence, it is important to study the truncated versions of these states.

In this section, we describe the truncated GKP states used in the \textit{Strawberry Fields} Python library. The exact finite-energy square-lattice GKP basis states are
\begin{equation}\label{eq:truncated_GKP}
\ket{k_{\epsilon}}\equiv \mathcal{N}_{\epsilon}\sum_{t=-z_{\mathrm{max}}}^{z_{\mathrm{max}}}c_t^{(k)}D(\alpha_t^{(k)})S(r_\epsilon)\ket{0},
\end{equation}
where $k\in\{0,1\}$ denotes the $0_L$ or $1_L$ respectively. The coefficients are given by
\begin{subequations}
\begin{align}
c_t^{(k)} &= \exp\left[-\frac{\pi}{2}\tanh(\epsilon) (k+2t)^2 \right],\\ \label{eq:alpha_t_k}
\alpha_t^{(k)} &= \sqrt{\frac{\pi}{2}} \frac{k+2t}{\cosh \epsilon}, \\
r_\epsilon &= -\frac{1}{2} \log(\tanh \epsilon), \\
\mathcal{N}_{\epsilon}^{-2} &= \sum_t \left| c_t^{(k)} \right|^2,
\end{align}
\end{subequations}
with $z_{\mathrm{max}}$ denoting the lattice cutoff, $t$ an integer representing the lattice index,
\begin{equation}
t = -z_{\mathrm{max}},...,0,...,z_{\mathrm{max}},
\end{equation}
and $\epsilon$ denoting the finite-energy parameter. In the limit $\epsilon\rightarrow 0$, we recover the ideal GKP state.

We can now find the phase error for the desired $T_{\mathrm{GKP}}$ gate:
\begin{subequations}\label{eq:delta_theta}
\begin{align}
\delta \theta(\epsilon) &\equiv \theta_{\varepsilon}-\frac{\pi}{4} \\
&= \arg M_{11}(\epsilon) - \arg M_{00}(\epsilon) - \frac{\pi}{4},
\end{align}
\end{subequations}
where the matrix elements $M_{ij}$ are
\begin{equation}\label{eq:M_t_gkp}
M_{ij}(\epsilon) \equiv \bra{i_\epsilon} T_{\mathrm{GKP}} \ket{j_{\epsilon}}
\end{equation}
for $i,j \in \{0,1\}$. Note that in the ideal case $\epsilon \rightarrow 0$, the phase error is zero: $\delta \theta \rightarrow 0$. In Appendix~\ref{app:truncated} we show how to derive the truncated GKP states from the ideal states, and give an approximate analytical bound for $\delta \theta$.

The process fidelity of this approximate gate with the ideal $T_{\mathrm{GKP}}$ gate is
\begin{equation}
F_{\mathrm{proc}} = \frac{|\mathrm{Tr}(T^{\dagger}_{\mathrm{GKP}}M)|}{4} = \cos^2\left(\frac{\delta \theta}{2}\right) ,
\end{equation}
where $M$ is the matrix with elements given by \cref{eq:M_t_gkp}.

\cref{fig:proc_infid_GKP} shows the process infidelity $1-F_{\mathrm{proc}}$ of the holonomic $T_{\mathrm{GKP}}$ gate as a function of the finite-energy parameter $\epsilon$. The blue markers are obtained using the Strawberry Fields library, while the orange curve is the smooth approximation obtained from the corrected finite-energy kernel model using \cref{eq:trunc_Airy}. Excellent agreement is observed throughout the entire range $0.05\leq\epsilon\leq0.40$, indicating that the kernel approximation captures the dominant finite-energy correction $\epsilon$; the smaller the energy correction, the smaller the infidelity, reaching a process infidelity of $1-F_{\mathrm{proc}} \approx 1.4\times 10^{-7}$ for $\epsilon = 0.05$. Our interpretation is that the finite-energy effects primarily act to renormalize the logical phase acquired by the $|1_L\rangle$ component, while inducing negligible logical-state mixing.

We now will calculate the \textit{leakage probability} due to measurement. First, we define the logical projector built from the truncated finite-energy basis:
\begin{equation}\label{eq:Proj_trunc_GKP}
\mathbb{P}_{\epsilon}\equiv W_{\epsilon}(W_{\epsilon}^{\dagger}W_{\epsilon})^{-1}W_{\epsilon}^{\dagger},
\end{equation}
where $W_{\epsilon}$ is a matrix whose columns are the truncated basis vectors in \cref{eq:truncated_GKP},
\begin{equation}
W_{\epsilon}\equiv \left[\,\ket{0_{\epsilon}}\, , \, \ket{1_{\epsilon}}\,\right],
\end{equation}
and the orthogonal projector is 
\begin{equation}
\mathbb{Q}_{\epsilon}= I - \mathbb{P}_{\epsilon}.
\end{equation}

We now consider a rotating-frame Lindblad master equation analogous to that in \cref{eq:Lindblad_RF_cat_code}, but substituting the unitary defined in \cref{eq:man_of_the_year} for the rotating unitary $V(t)$ and the truncated projector in \cref{eq:Proj_trunc_GKP} for $\mathbb{P}_0$. From this, we get an analytical bound on the leakage probability:
\begin{equation}\label{eq:bound_leak_GKP}
p_{\mathrm{leak}} \leq \frac{4}{\kappa t_f} \int_{0}^1 ds\, ||\mathbb{Q}_{\epsilon}V^{\dagger}(s)\dot{V}(s)\mathbb{P}_{\epsilon}|| .
\end{equation}

\begin{figure}
\centering
\includegraphics[width=\linewidth]{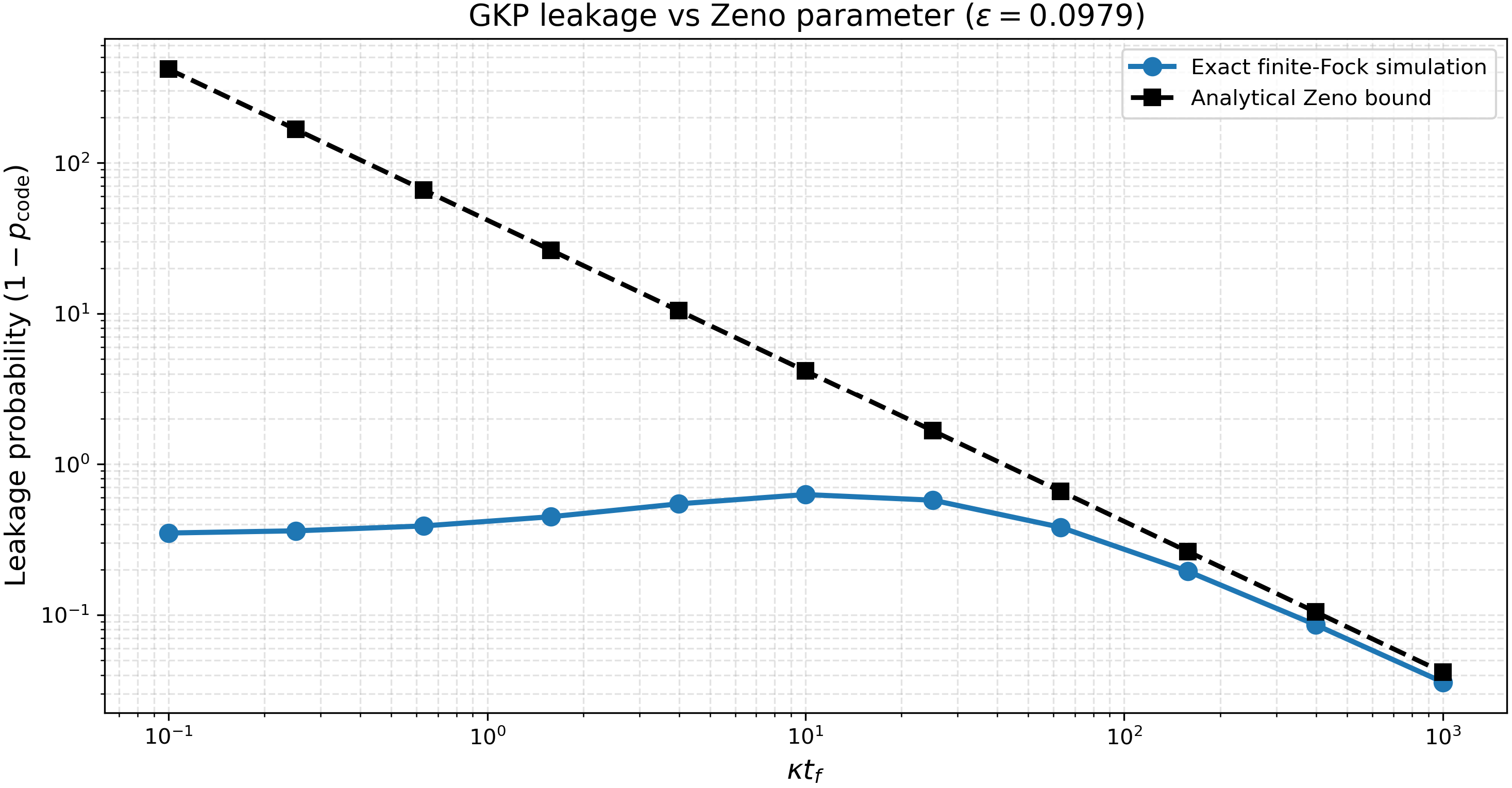}
\caption{Leakage probability $1-p_{\rm code}$ as a function of the Zeno parameter $\kappa t_f$ for the finite-energy GKP code with $\epsilon=0.0979$. Blue markers show the numerical finite-Fock simulation, while the dashed black curve corresponds to the analytical Zeno bound.
}
\label{fig:pleak_GKP}
\end{figure}

\cref{fig:pleak_GKP} shows the leakage probability $p_{\rm leak}$ as a function of the Zeno parameter $\kappa t_f$. As the measurement strength increases, leakage is progressively suppressed, demonstrating the onset of the quantum Zeno regime. The analytical bound correctly captures the asymptotic scaling $1/(\kappa t_f)$ and remains an upper bound throughout the explored parameter range. The bound becomes increasingly tight in the strong-measurement regime, where the system dynamics are effectively confined to the finite-energy GKP codespace.

\section{Conclusion}

We have extended continuous measurement-based holonomic quantum computation (CMHQC) to bosonic quantum error-correcting codes and presented explicit implementations for both four-component cat codes and Gottesman–Kitaev–Preskill (GKP) codes. In this framework, logical operations arise from the geometric evolution of the continuously monitored, time-dependent codespace, while the quantum Zeno effect suppresses departures from the instantaneous encoded subspace.

For cat codes, we introduced a family of squeezed-cat trajectories whose projected Wilczek–Zee connection can generate arbitrary logical $Z_L$ rotations, including the non-Clifford $T_L$ gate. For GKP codes, we constructed a translated-lattice trajectory whose associated holonomy reproduces the logical $T_{\mathrm{GKP}}$ gate through a purely geometric mechanism. In both cases, the desired logical operation emerges from the geometry of the measurement path rather than from direct Hamiltonian control.

Going beyond the gate constructions themselves, we analyzed the error-correcting properties of the instantaneous codespaces by deriving dressed Knill–Laflamme conditions for the relevant bosonic error models. We further investigated the effects of finite measurement strength and obtained analytical bounds on leakage out of the encoded subspace. These results demonstrate that the protocols recover the ideal holonomic evolution in the strong-measurement regime while maintaining the underlying error-correction structure of the codes.

Our work establishes bosonic codes as a natural platform for continuous measurement-based holonomic control and provides a concrete route toward fault-tolerant geometric quantum computation without requiring direct implementation of logical Hamiltonians. More broadly, it connects geometric phases, continuous quantum measurement, and bosonic quantum error correction within a unified framework. Directions for future research include extending these methods to other bosonic encodings, incorporating realistic finite-energy code states and experimental imperfections, and developing multi-qubit holonomic gates capable of supporting a complete fault-tolerant universal gate set.

%\newpage
\acknowledgments
JGN, AL and TAB would like to thank Onkar Apte, Andreas Bauer, Christopher Gerhard, Miles Gorman, Yucheng He, Daniel Lidar, Seth Lloyd, Arjun Sathyamoorthy, Yuanjia Wang, Shixin Wu and Dawei Zhong for interesting and useful discussions. This work was supported in part by the U.S. Army Research Laboratory and the U.S. Army Research Office under contract/grant number W911NF2310255, and by NSF Grant FET-2316713.

\bibliography{references}

@article{zanardi_1999,
title = {Holonomic quantum computation},
journal = {Physics Letters A},
volume = {264},
number = {2},
pages = {94-99},
year = {1999},
issn = {0375-9601},
doi = {https://doi.org/10.1016/S0375-9601(99)00803-8},
url = {https://www.sciencedirect.com/science/article/pii/S0375960199008038},
author = {Paolo Zanardi and Mario Rasetti}
}

@article{Mommers_2022,
  title = {Universal quantum computation and quantum error correction using discrete holonomies},
  author = {Mommers, Cornelis J. G. and Sj\"oqvist, Erik},
  journal = {Phys. Rev. A},
  volume = {105},
  issue = {2},
  pages = {022402},
  numpages = {6},
  year = {2022},
  month = {Feb},
  publisher = {American Physical Society},
  doi = {10.1103/PhysRevA.105.022402},
  url = {https://link.aps.org/doi/10.1103/PhysRevA.105.022402}
}

@article{Wilczek_Zee,
  title = {Appearance of Gauge Structure in Simple Dynamical Systems},
  author = {Wilczek, Frank and Zee, A.},
  journal = {Phys. Rev. Lett.},
  volume = {52},
  issue = {24},
  pages = {2111--2114},
  numpages = {0},
  year = {1984},
  month = {Jun},
  publisher = {American Physical Society},
  doi = {10.1103/PhysRevLett.52.2111},
  url = {https://link.aps.org/doi/10.1103/PhysRevLett.52.2111}
}

@article{Lanka2025,
  author       = {Lanka, Anirudh and Garcia-Nila, Juan and Brun, Todd A.},
  title        = {Continuous Measurement-Based Holonomic Quantum Computation},
  journal      = {arXiv preprint},
  eprint       = {2510.06725},
  archivePrefix = {arXiv},
  year         = {2025},
  primaryClass = {quant-ph},
  url          = {https://arxiv.org/abs/2510.06725}
}

@article{Lanka2026,
  author       = {Lanka, Anirudh and Garcia-Nila, Juan and Brun, Todd A.},
  title        = {Steering Paths Mid-Flight for Fault-Tolerance in Measurement-Based Holonomic Gates},
  journal      = {arXiv preprint},
  eprint       = {2603.02552v1},
  archivePrefix = {arXiv},
  year         = {2026},
  primaryClass = {quant-ph},
  url          = {https://arxiv.org/abs/2603.02552}
}

@article{Mirrahimi_2014,
doi = {10.1088/1367-2630/16/4/045014},
url = {https://doi.org/10.1088/1367-2630/16/4/045014},
year = {2014},
month = {apr},
publisher = {IOP Publishing},
volume = {16},
number = {4},
pages = {045014},
author = {Mirrahimi, Mazyar and Leghtas, Zaki and Albert, Victor V and Touzard, Steven and Schoelkopf, Robert J and Jiang, Liang and Devoret, Michel H},
title = {Dynamically protected cat-qubits: a new paradigm for universal quantum computation},
journal = {New Journal of Physics}
}

@article{Nguyen2026,
  title = {Fault-tolerant non-Clifford Gottesman-Kitaev-Preskill gates using polynomial phase gates and on-demand noise biasing},
  author = {Nguyen, Minh T. P. and Shaw, Mackenzie H.},
  journal = {Phys. Rev. Res.},
  volume = {8},
  issue = {3},
  pages = {033107},
  numpages = {39},
  year = {2026},
  month = {Jul},
  publisher = {American Physical Society},
  doi = {10.1103/2pg9-nsn8},
  url = {https://link.aps.org/doi/10.1103/2pg9-nsn8}
}

@article{
Leghtas_2015,author = {Z. Leghtas  and S. Touzard  and I. M. Pop  and A. Kou  and B. Vlastakis  and A. Petrenko  and K. M. Sliwa  and A. Narla  and S. Shankar  and M. J. Hatridge  and M. Reagor  and L. Frunzio  and R. J. Schoelkopf  and M. Mirrahimi  and M. H. Devoret },
title = {Confining the state of light to a quantum manifold by engineered two-photon loss},
journal = {Science},
volume = {347},
number = {6224},
pages = {853-857},
year = {2015},
doi = {10.1126/science.aaa2085},
URL = {https://www.science.org/doi/abs/10.1126/science.aaa2085},
eprint = {https://www.science.org/doi/pdf/10.1126/science.aaa2085}}

@article{Touzard_2018,
  title = {Coherent Oscillations inside a Quantum Manifold Stabilized by Dissipation},
  author = {Touzard, S. and Grimm, A. and Leghtas, Z. and Mundhada, S. O. and Reinhold, P. and Axline, C. and Reagor, M. and Chou, K. and Blumoff, J. and Sliwa, K. M. and Shankar, S. and Frunzio, L. and Schoelkopf, R. J. and Mirrahimi, M. and Devoret, M. H.},
  journal = {Phys. Rev. X},
  volume = {8},
  issue = {2},
  pages = {021005},
  numpages = {7},
  year = {2018},
  month = {Apr},
  publisher = {American Physical Society},
  doi = {10.1103/PhysRevX.8.021005},
  url = {https://link.aps.org/doi/10.1103/PhysRevX.8.021005}
}

@article{Heeres2015,
  title = {Cavity State Manipulation Using Photon-Number Selective Phase Gates},
  author = {Heeres, Reinier W. and Vlastakis, Brian and Holland, Eric and Krastanov, Stefan and Albert, Victor V. and Frunzio, Luigi and Jiang, Liang and Schoelkopf, Robert J.},
  journal = {Phys. Rev. Lett.},
  volume = {115},
  issue = {13},
  pages = {137002},
  numpages = {5},
  year = {2015},
  month = {Sep},
  publisher = {American Physical Society},
  doi = {10.1103/PhysRevLett.115.137002},
  url = {https://link.aps.org/doi/10.1103/PhysRevLett.115.137002}
}

@article{Krastanov_2015,
  title = {Universal control of an oscillator with dispersive coupling to a qubit},
  author = {Krastanov, Stefan and Albert, Victor V. and Shen, Chao and Zou, Chang-Ling and Heeres, Reinier W. and Vlastakis, Brian and Schoelkopf, Robert J. and Jiang, Liang},
  journal = {Phys. Rev. A},
  volume = {92},
  issue = {4},
  pages = {040303(R)},
  numpages = {5},
  year = {2015},
  month = {Oct},
  publisher = {American Physical Society},
  doi = {10.1103/PhysRevA.92.040303},
  url = {https://link.aps.org/doi/10.1103/PhysRevA.92.040303}
}

@article{Grimm_2020,
   title={Stabilization and operation of a Kerr-cat qubit},
   volume={584},
   ISSN={1476-4687},
   url={http://dx.doi.org/10.1038/s41586-020-2587-z},
   DOI={10.1038/s41586-020-2587-z},
   number={7820},
   journal={Nature},
   publisher={Springer Science and Business Media LLC},
   author={Grimm, A. and Frattini, N. E. and Puri, S. and Mundhada, S. O. and Touzard, S. and Mirrahimi, M. and Girvin, S. M. and Shankar, S. and Devoret, M. H.},
   year={2020},
   month=Aug, pages={205–209} }

@article{Kang_2022,
  title = {Nonadiabatic geometric quantum computation with cat-state qubits via invariant-based reverse engineering},
  author = {Kang, Yi-Hao and Chen, Ye-Hong and Wang, Xin and Song, Jie and Xia, Yan and Miranowicz, Adam and Zheng, Shi-Biao and Nori, Franco},
  journal = {Phys. Rev. Res.},
  volume = {4},
  issue = {1},
  pages = {013233},
  numpages = {16},
  year = {2022},
  month = {Mar},
  publisher = {American Physical Society},
  doi = {10.1103/PhysRevResearch.4.013233},
  url = {https://link.aps.org/doi/10.1103/PhysRevResearch.4.013233}
}

@article{Puri_2020,
   title={Bias-preserving gates with stabilized cat qubits},
   volume={6},
   ISSN={2375-2548},
   url={http://dx.doi.org/10.1126/sciadv.aay5901},
   DOI={10.1126/sciadv.aay5901},
   number={34},
   journal={Science Advances},
   publisher={American Association for the Advancement of Science (AAAS)},
   author={Puri, Shruti and St-Jean, Lucas and Gross, Jonathan A. and Grimm, Alexander and Frattini, Nicholas E. and Iyer, Pavithran S. and Krishna, Anirudh and Touzard, Steven and Jiang, Liang and Blais, Alexandre and Flammia, Steven T. and Girvin, S. M.},
   year={2020},
   month=Aug }

@article{Zhang_2024,
   title={Non-adiabatic holonomic quantum operations in continuous variable systems},
   volume={67},
   ISSN={1869-1927},
   url={http://dx.doi.org/10.1007/s11433-023-2339-x},
   DOI={10.1007/s11433-023-2339-x},
   number={6},
   journal={Science China Physics, Mechanics \& Astronomy},
   publisher={Springer Science and Business Media LLC},
   author={Zhang, Hao-Long and Kang, Yi-Hao and Wu, Fan and Yang, Zhen-Biao and Zheng, Shi-Biao},
   year={2024},
   month=May }

@article{GKP2001,
  title = {Encoding a qubit in an oscillator},
  author = {Gottesman, Daniel and Kitaev, Alexei and Preskill, John},
  journal = {Phys. Rev. A},
  volume = {64},
  issue = {1},
  pages = {012310},
  numpages = {21},
  year = {2001},
  month = {Jun},
  publisher = {American Physical Society},
  doi = {10.1103/PhysRevA.64.012310},
  url = {https://link.aps.org/doi/10.1103/PhysRevA.64.012310}
}

@article{Glancy_2006,
   title={Error analysis for encoding a qubit in an oscillator},
   volume={73},
   ISSN={1094-1622},
   url={http://dx.doi.org/10.1103/PhysRevA.73.012325},
   DOI={10.1103/physreva.73.012325},
   number={1},
   journal={Physical Review A},
   publisher={American Physical Society (APS)},
   author={Glancy, S. and Knill, E.},
   year={2006},
   month=Jan }

@article{Fukui2018,
  author = {Fukui, Kosuke and Tomita, Akihisa and Okamoto, Atsushi and Fujii, Keisuke},
  title = {High-threshold fault-tolerant quantum computation with analog quantum error correction},
  journal = {Phys. Rev. X},
  volume = {8},
  pages = {021054},
  year = {2018}
}

@article{Fluhmann2019,
  author = {Flühmann, Christa and Nguyen, Timothy and Marinelli, Massimo and Negnevitsky, Vlad and Mehta, Karan and Home, Jonathan},
  title = {Encoding a qubit in a trapped-ion mechanical oscillator},
  journal = {Nature},
  volume = {566},
  pages = {513--517},
  year = {2019}
}

@article{deNeeve2025,
  author = {de Neeve, B. et al.},
  title = {Error correction with the Gottesman-Kitaev-Preskill code},
  journal = {Nature},
  year = {2025}
}

@article{Albert_2016,
  title = {Holonomic Quantum Control with Continuous Variable Systems},
  author = {Albert, Victor V. and Shu, Chi and Krastanov, Stefan and Shen, Chao and Liu, Ren-Bao and Yang, Zhen-Biao and Schoelkopf, Robert J. and Mirrahimi, Mazyar and Devoret, Michel H. and Jiang, Liang},
  journal = {Phys. Rev. Lett.},
  volume = {116},
  issue = {14},
  pages = {140502},
  numpages = {6},
  year = {2016},
  month = {Apr},
  publisher = {American Physical Society},
  doi = {10.1103/PhysRevLett.116.140502},
  url = {https://link.aps.org/doi/10.1103/PhysRevLett.116.140502}
}

@article{Menicucci_2014,
  title = {Fault-Tolerant Measurement-Based Quantum Computing with Continuous-Variable Cluster States},
  author = {Menicucci, Nicolas C.},
  journal = {Phys. Rev. Lett.},
  volume = {112},
  issue = {12},
  pages = {120504},
  numpages = {5},
  year = {2014},
  month = {Mar},
  publisher = {American Physical Society},
  doi = {10.1103/PhysRevLett.112.120504},
  url = {https://link.aps.org/doi/10.1103/PhysRevLett.112.120504}
}
\appendix

\section{Measurement-based holonomic quantum computing}\label{app:CMBHQC}

The natural geometric object is the complex Grassmannian manifold of rank-$2$ projections in the set of all bounded linear operators on the Hilbert space
$\mathcal H_{\mathrm{osc}}$:
\begin{equation}
\mathbf{Gr_2} = \{ \mathbb{P} \in \mathcal B(\mathcal H_{\mathrm{osc}}) \;|\; \mathbb{P}^2 = \mathbb{P},\; \mathbb{P}^\dagger = \mathbb{P},\; \mathrm{rank}(\mathbb{P})=2 \},
\end{equation}
where each $\mathbb{P} \in \mathbf{Gr_2}$ represents a two-dimensional subspace of $\mathcal H_{\mathrm{osc}}$. The set of basis vectors is not unique; hence, different bases for the same subspace are related through unitaries in the group $U(2)$.

The associated Stiefel manifold is
\begin{equation}
\mathcal F = \{ L \in \mathrm{Hom}\!\left(\mathbb{C}^2,\mathcal{H}_{\mathrm{osc}}\right) \;|\; L^\dagger L = I_2 \},
\end{equation}
where $L$ is an infinite-dimensional rectangular operator of rank 2 and the columns of $L$ form an orthonormal basis of the subspace. Geometrically, each point on the Grassmannian manifold has a \textit{fiber} comprising all possible orthonormal bases of the subspace, and the Stiefel manifold is the collection of all of them, a \textit{fiber bundle}.

The projection map $\pi:\mathcal{F}\rightarrow \mathbf{Gr_2}$ is defined as
\begin{equation}
\mathbb{P}\equiv L L^\dagger,
\end{equation}
and relates the two manifolds. Furthermore, any right action of $U(2)$ on $\mathcal F$ leaves the projector invariant, i.e., given $h \in U(2)$, it satisfies
\begin{equation}\label{eq:Andy}
(L,h) \mapsto Lh, \qquad \pi(Lh) = \pi(L).
\end{equation}
The canonical connection on $\mathcal F$ is given by
\begin{equation}\label{eq:connection}
\mathcal{A} = L^\dagger dL,
\end{equation}
which under a gauge transformation $L \to Lh$ becomes
\begin{equation}
\mathcal{A} \mapsto h^\dagger A h + h^\dagger dh.
\end{equation}

On the other hand, a unitary operator $V \in \mathcal U(\mathcal H_{\mathrm{osc}})$ acts on the elements of the Grassmannian manifold as
\begin{equation}\label{eq:Miranda}
\mathbb{P} \mapsto V \mathbb{P} V^\dagger.
\end{equation}
In our paper, for each value of the time parameter $t$, we choose the unitary $V(t)$ such that the projector $V \mathbb{P}_0 V^\dagger$ onto the instantaneous codespace is frequently measured, maintaining the state in the code space at all times. It satisfies
\begin{equation}
\mathbb{P}(t)=L(t)L^{\dagger}(t),
\end{equation}
and by the previous properties in \cref{eq:Miranda,eq:Andy,eq:projector_t} we have a general solution for the curve $L(t)$,
\begin{equation}\label{eq:L_of_t}
L(t)=V(t)L(0)h(t),  
\end{equation}
where $L(t)$ relates the state $\ket{\psi(t)}\in \mathcal{H}_{\text{osc}}$ to the reduced encoded state $\ket{\varphi(t)}\in \mathbb{C}^2$:
\begin{equation}\label{eq:psi_of_t}
\ket{\psi(t)}=L(t)\ket{\varphi(t)} .
\end{equation}

By the Schr\"odinger equation, the reduced dynamics satisfies
\begin{equation}
\frac{d}{dt}|\varphi(t)\rangle + L^\dagger \dot L |\varphi(t)\rangle = 0,
\end{equation}
whose solution is a time ordered exponential:
\begin{equation}
|\varphi(t)\rangle = \mathcal P \exp\!\left(-\int_0^t L^\dagger \dot L \, dt \right) |\varphi(0)\rangle.
\end{equation}
To eliminate the path dependence, we impose the {\it horizontal condition}, where the connection $\mathcal{A}$ in \cref{eq:connection} becomes zero (also known as the {\it parallel transport} equation):
\begin{equation}\label{eq:horizontal_cond}
L^{\dagger} \frac{dL}{dt}=0 \qquad \forall t.
\end{equation}

If the horizontal condition is imposed, we substitute $L(t)$ from \cref{eq:L_of_t} into \cref{eq:horizontal_cond} to obtain an equation for $h(t)\in U(2)$:
\begin{equation}
\frac{dh(t)}{dt} = - L^\dagger(0) V^\dagger(t)\frac{dV(t)}{dt} L(0) \, h(t),
\end{equation}
with solution
\begin{equation}\label{eq:h_oft}
h(t) = \mathcal P \exp\!\left(- \int_0^t L^\dagger(0) V^\dagger(t) \dot V(t) L(0) \, dt \right).
\end{equation}
Finally, combining all of the ingredients, since the initial reduced state can be written as $\ket{\varphi(0)} = L^{\dagger}(0) \ket{\psi(0)}$ from \cref{eq:psi_of_t}, and using the general expression for the horizontal lift $L(t)$ in \cref{eq:L_of_t}, we obtain an equation for the dynamics of a code state:
\begin{subequations}\label{eq:psi_of_t}
\begin{align}
\ket{\psi(t)} &= L(t)L^{\dagger}(0) \ket{\psi(0)} \\
&= V(t) L(0) h(t) L^{\dagger}(0) \ket{\psi(0)},
\end{align} 
\end{subequations}
where $h(t)$ is given in \cref{eq:h_oft}.

\section{Displacement operator}\label{app:displace}

A useful property relating the annihilation operator $a$ and creation operator $a^{\dagger}$ is their commutation relation,
\begin{equation}\label{eq:comm_a}
[ a,a^\dagger ] = I.
\end{equation}
Let us state some useful properties of the displacement operators defined in \cref{eq:displacement}, with $\alpha$ and $\beta$ two arbitrary complex numbers.

First, there are unitary conditions
\begin{subequations}
\begin{align}
D^{\dagger}(\alpha) &= D(-\alpha) , \\
D(\alpha) D^{\dagger}(\alpha) &= I ,
\end{align}
\end{subequations}
along with the displacement property 
\begin{subequations}\label{eq:displace_prop}
\begin{align}
D^{\dagger}(\alpha) a D(\alpha) &= a + \alpha I , \\
D(\alpha) a D^{\dagger}(\alpha) &= a - \alpha I ,
\end{align}
\end{subequations}
These properties can be proven using the commutation relations of ladder operators in \cref{eq:comm_a}.

From the Baker–Campbell–Hausdorff (BCH) formula:
\begin{subequations}\label{eq:BCH_Disp}
\begin{align}\label{eq:YL_Disp}
D(\alpha) D(\beta) &= e^{(\alpha\beta^* - \alpha^* \beta)/2} D(\alpha+\beta), \\ \label{eq:Comm_Disp}
D(\alpha)D(\beta) &= e^{\alpha\beta^* - \alpha^* \beta} D(\beta) D(\alpha).
\end{align}
\end{subequations}

The coherent states obey
\begin{equation}
\braket{\beta|\gamma} = \exp\left(-\frac{|\beta|^2}{2} - \frac{|\gamma|^2}{2} + \beta^* \gamma \right).
\end{equation}

\section{Holonomic protocol for cat codes}\label{app:WZ_app}

\subsection{Horizontal lift for cat codes}\label{app:h_t_cat}

The dynamics of a state $\ket{\psi(t)}$ in the cat-qubit codespace is given in \cref{eq:psi_of_t}, where the horizontal lift $L(t)$ is given in \cref{eq:L_of_t} and it satisfies the parallel transport condition \cref{eq:horizontal_cond} \cite{Wilczek_Zee}. 

To obtain the unitary $h(t)$ in \cref{eq:h_oft} for the cat code described in \cref{sec:cat_code}, we use the rotating unitary in \cref{eq:rotating_V}:
\begin{equation}\label{eq:connection_WZ_cat}
V^{\dagger}(t)\dot{V}(t) = i\dot{\phi}(t) S^{\dagger}(\xi(t)) Z_L S(\xi(t)) + S^{\dagger}(\xi(t)) \dot{S}(\xi(t)),
\end{equation}
where to reduce the first term we use the property
\begin{equation}\label{eq:prop_N_a}
e^{i\theta N} a e^{-i \theta N} = e^{-i\theta} a,
\end{equation}
which can be proven using the commutator of the annihilation operator $a$ with the number operator $N$ in \cref{eq:N_op}, $[N,a]=-a$. We define the auxiliary function $f_N$
\begin{equation}
f_N(\theta)\equiv e^{i\theta N} a e^{-i \theta N}, \\
\end{equation}
which satisfies the differential equation
\begin{subequations}
\begin{align}
\frac{df_N}{d\theta} = i e^{i\theta N} [N,a] e^{-i\theta N} &= i e^{i\theta N} (-a) e^{-i\theta N} = -i f_N(\theta) \\
\implies f_N(\theta) = e^{-i\theta} f_N(0) &= e^{-i\theta} a,
\end{align}
\end{subequations}
where the last implication is the result \cref{eq:prop_N_a}. If we square it we obtain
\begin{equation}
e^{i\theta N} a^2 e^{-i\theta N} = e^{-2i\theta}a^2.
\end{equation}
A similar result can be proven for $a^{\dagger \, 2}$. In particular, for $\theta=\frac{\pi}{2}$ and $Z_L$ as defined in \cref{eq:ZL_cat_op}, we get
\begin{subequations}
\begin{align}
Z_L a^2 Z_L^{\dagger} &= e^{-i\pi}a^2 = -a^2, \\
Z_L (\xi^* a^2 - \xi a^{\dagger \, 2}) Z_L^{\dagger} &= -(\xi^* a^2 - \xi a^{\dagger \, 2}).
\end{align} 
\end{subequations}
Exponentiating the last expression leads to the result
\begin{subequations}
\begin{align}
Z_LS(\xi) = S^{\dagger}(\xi) Z_L &= S(-\xi) Z_L, \\
S^{\dagger}(\xi) Z_L S(\xi) &= S(-2\xi) Z_L.
\end{align}
\end{subequations}
Now using a similar property in \cref{eq:XL_cat}, we obtain
\begin{equation}
\mathbb{P}_0(-\xi^*a^2+\xi a^{\dagger \, 2})\mathbb{P}_0=\mathbb{P}_0(-\xi^*\alpha^2+\xi \alpha^{\dagger \, 2})X_L \mathbb{P}_0+\mathcal{O}(e^{-2|\alpha|^2}),
\end{equation}
where taking the exponential, assuming leakage terms are negligible at large $|\alpha|$, we obtain
\begin{align}\label{eq:final_h_cat}
\mathbb{P}_0 S(-2\xi) \mathbb{P}_0 =& \mathbb{P}_0 \exp\left[-(\xi^*\alpha^2-\xi \alpha^{* 2}) X_L\right] \mathbb{P}_0 \\
&+ \mathcal{O}(e^{-|\alpha|^2}).
\end{align}
So we obtain the projected first term in \cref{eq:connection_WZ_cat},
\begin{subequations}
\begin{align}
\mathbb{P}_0 S(-2\xi) Z_L \mathbb{P}_0 &= \mathbb{P}_0 S(-2\xi) \mathbb{P}_0 Z_L \\
&\approx \mathbb{P}_0 e^{-(\xi^*\alpha^2 - \xi \alpha^{* 2}) X_L} Z_L \mathbb{P}_0,
\end{align}
\end{subequations}
and, assuming the phase $\vartheta$ of $\xi$ is constant in time, the second term in \cref{eq:connection_WZ_cat} is
\begin{equation}
S^{\dagger}(\xi(t)) \dot{S}(\xi(t)) = -\frac{\dot{r}(t)}{2} (e^{i\vartheta} a^{\dagger \, 2} - e^{-i\vartheta} a^{2}).
\end{equation}

Now we will analyze these two terms, using the amplitudes and phases of $\alpha = |\alpha|e^{i\varphi}$ and $\xi(t) = r(t)e^{i\vartheta(t)}$ in \cref{eq:xi}, we define the exponential argument
\begin{equation}
\chi(t) \equiv -i(\xi^*\alpha^2-\xi \alpha^{* 2}) = 2\, |\alpha|^2\, r(t) \sin[2\varphi - \vartheta(t)].
\end{equation}
So the projected Wilczek connection in \cref{eq:connection_WZ_cat} becomes, in the large $|\alpha|^2$ approximation,
\begin{equation}
\mathbb{P}_0 V^{\dagger} \dot{V} \mathbb{P}_0 \approx i\dot{\phi}(t) \left[\cos\chi(t)\, Z_L - \sin\chi(t)\, Y_L\right] + \frac{i}{2} \dot{\chi}(t) X_L.
\end{equation}
To make the unitary $h(t)$ in \cref{eq:h_oft} trivial, $h(t)\approx I$,  we choose a time-dependent amplitude and phase choice in \cref{eq:xi} for the horizontal lift $L(t)$:
\begin{subequations}
\begin{align}
\chi(t) = 2|\alpha|^2 r(t) \sin \frac{\pi}{2} &= j_{0,1} \sin\frac{2\pi t}{t_f}, \\
\int_{0}^{t_f} \cos\chi(t) dt &= t_f J_0(j_{0,1}) = 0,\\
\int_{0}^{t_f} \sin\chi(t)dt &= 0.
\end{align}
\end{subequations}

\subsection{Error correction conditions}\label{app:KL_cat_code}

To find the Knill-Laflamme conditions for the cat code, we start with the parity from \cref{eq:parity_op} and annihilation operator from \cref{eq:coherent}, whose anticommutation property gives
\begin{equation}\label{eq:Pi_a}
\Pi a \Pi=-a,
\end{equation}
which follows immediately from \cref{eq:prop_N_a} with $\theta=\pi$.

Using the fact that the parity operator is a stabilizer, we see that it commutes with the codespace projector from \cref{eq:cat_proj}, and can be absorbed into that projector:
\begin{equation}\label{eq:P0_Pi}
    \Pi \mathbb{P}_0 = \mathbb{P}_0 \Pi = \mathbb{P}_0.
\end{equation}
The property in \cref{eq:a_corr_cat} then follows from \cref{eq:Pi_a}:
\begin{equation}
\mathbb{P}_0 a \mathbb{P}_0 = \Pi \mathbb{P}_0 a \mathbb{P}_0 \Pi =  \mathbb{P}_0 \Pi a \Pi \mathbb{P}_0 = -\mathbb{P}_0 a \mathbb{P}_0.
\end{equation}
The property in \cref{eq:a_dag_corr_cat} follows similarly from
\begin{equation}\label{eq:Pi_a_dag}
\Pi a^{\dagger} \Pi = -a^{\dagger},
\end{equation}
which follows immediately from conjugating \cref{eq:Pi_a}. \cref{eq:N_cat} then follows from $N = a^{\dagger}a$ and the commutation of number and parity operators:
\begin{equation}
a^{\dagger} a \mathbb{P}_0 = N \mathbb{P}_0 = \frac{1}{2} N (I+\Pi) \mathbb{P}_{\text{4cat}} = \mathbb{P}_0 N.
\end{equation}
The property in \cref{eq:Nplus1_cat} follows from \cref{eq:comm_a}:
\begin{equation}
aa^{\dagger} = a^{\dagger}a + I = N + I.
\end{equation}
Now, using the property of the annihilation operator over the cat states from \cref{eq:a_over_cat}, we obtain the expectation value of \cref{eq:N_op}:
\begin{equation}
\bra{C_{\alpha}^k} N \ket{C_{\alpha}^k} = \bra{C_{\alpha}^k} a^{\dagger}a \ket{C_{\alpha}^k} = |\alpha|^2 + \mathcal{O}(e^{-|\alpha|^2}).
\end{equation}
Hence, the properties in \cref{eq:KL_Nop_cat} follow.

Since the parity operator in \cref{eq:parity_op} is a function of the number operator and commutes with both the squeezing operator in \cref{eq:squeezing_op} and the logical $Z_L$ operator in \cref{eq:ZL_cat_op}, we have
\begin{equation}
    [\Pi, S(\xi)]=0\qquad [\Pi, Z_L]=0.
\end{equation}
Then, using the dressed operator definition in \cref{eq:a_of_t}, the Knill-Laflamme conditions for the time-dependent projector reduce to 
\begin{subequations}
\begin{align}
\mathbb{P}_0 a(t) \mathbb{P}_0 &= \mathbb{P}_0 S^{\dagger}(\xi) e^{-i\phi Z_L} a e^{i\phi Z_L} S(\xi) \mathbb{P}_0 \\
&= \mathbb{P}_0 \Pi S^{\dagger}(\xi) e^{-i\phi Z_L} a e^{i\phi Z_L} S(\xi) \Pi \mathbb{P}_0 \\
&= \mathbb{P}_0 S^{\dagger}(\xi) e^{-i\phi Z_L} \Pi a\Pi e^{i\phi Z_L} S(\xi) \mathbb{P}_0 \\
&= -\mathbb{P}_0 a(t) \mathbb{P}_0,
\end{align}
\end{subequations}
where in the second line we used \cref{eq:P0_Pi} and in the third line \cref{eq:Pi_a}. So \cref{eq:KL_Pt_a} follows from the last equality by multiplying by $V(t)$ from the left and by $V^{\dagger}(t)$ from the right. Similarly, the time-dependent $a^{\dagger}(t)$ in \cref{eq:a_dagger_of_t} also obeys the Knill-Laflamme condition:
\begin{equation}
\mathbb{P}_0 a^\dagger(t) \mathbb{P}_0 = 0.
\end{equation}

The property in \cref{eq:KL_adag_elxokas} follows similarly from \cref{eq:Pi_a_dag}, while the property \cref{eq:KL_N_Eleazar} follows from
\begin{align}
N(t) =& S^{\dagger}(\xi(t)) N S(\xi(t)) = \cosh(2r)N + \sinh^2(r)I \nonumber\\
&-\frac{1}{2} \sinh(2r) (e^{i\vartheta} a^{\dagger\, 2} + e^{-i\vartheta} a^{ 2}),
\end{align}
where we used the expression $\xi = re^{i\vartheta}$ in \cref{eq:xi}
and the property over the squeezed annihilation operator
\begin{equation}
S^{\dagger}(\xi) a S(\xi) = a \cosh r -e^{i\vartheta} a^{\dagger} \sinh r.
\end{equation}
If we compute the projected time-dependent operator in the large cat approximation we get
%\begin{subequations}
\begin{align}
&\mathbb{P}_0 N(t) \mathbb{P}_0 = \left[\cosh(2r) |\alpha|^2+\sinh^2(r)\right] \mathbb{P}_0 \nonumber \\
&-\frac{1}{2} \sinh(2r) (e^{i\vartheta} \alpha^{*\, 2} + e^{-i\vartheta} \alpha^{2}) X_L \mathbb{P}_0 + \mathcal{O}(e^{-|\alpha|^2}) \nonumber\\
&= \left[\cosh(2r) |\alpha|^2 + \sinh^2(r) \right] \mathbb{P}_0 + \mathcal{O}(e^{-|\alpha|^2}),
\end{align}
%\end{subequations}
where in the last equality we used the choice of the squeezed phase in \cref{eq:xi}:
\begin{equation}
e^{i\vartheta} \alpha^{*\, 2} + e^{-i\vartheta} \alpha^{ 2} = 2|\alpha|^2 \cos(\vartheta-2\varphi) = 0.
\end{equation}

\subsection{Leakage probability due to measurement}\label{app:Leak_prob_cat}

Now we discuss how to derive the leakage probability in \cref{eq:leak_prob_cat} using the Lindblad ensemble average equation in its rotated-frame version \cref{eq:Lindblad_RF_cat_code}. 

%\label{eq:B_ortho}

We have defined the leakage subspace $\mathbb{Q}_0$ in \cref{eq:Q0_ortho}, such that we decompose the whole Hilbert space as in \cref{eq:H_C_Cperp}. In this basis, we rewrite the state as a density matrix in $2\times 2$ block form, for an initial logical basis state $\ket{j_L}$, $j=0,1$, such that we keep the leading code and leakage amplitudes:
\begin{equation}
\tilde{\rho}_j(t) = \begin{pmatrix}
p_j(t) & c_j(t) \\
c_j^*(t) & q_j(t) \\
\end{pmatrix},
\end{equation}
and we define the effective anti-Hermitian operator, which is proportional to the effective rotating-frame Hamiltonian, in \cref{eq:Lindblad_RF_cat_code}
\begin{equation}
K(t)\equiv V^{\dagger}(t)\dot{V}(t) ,
\end{equation}
where the matrix elements are
%\begin{subequations}
\begin{align}
K_{jj} &\equiv \bra{j_L}K\ket{j_L}, \\ 
K_{j,\perp} &\equiv \bra{j_L}B\ket{j_L^{\perp}}, \nonumber
\end{align}
%\end{subequations}
where $B$ is given in \cref{eq:B_ortho} and $\{j_L^{\perp}\}$ works as an orthonormal basis for the leakage subspace. Furthermore, let us define
\begin{equation}
z_j(t)\equiv p_j(t)-q_j(t).
\end{equation}

We solve the rotating-frame master equation in \cref{eq:Lindblad_RF_cat_code}, whose solutions for the population and coherences are
%\begin{subequations}
\begin{align}\label{eq:c_eq_dif}
\dot{p}_j &= -(K_{j,\perp}c_j^* + K_{j,\perp}^*c_j), \nonumber\\
\dot{q}_j &= +(K_{j,\perp}c_j^* + K_{j,\perp}^*c_j), \\
\dot{c}_j &= -2K_{jj} c_j - \frac{\kappa}{2} c_j + K_{j,\perp}(p_j-q_j). \nonumber
\end{align}
%\end{subequations}
Therefore, by using the exponential trick we define new variables
\begin{subequations}\label{eq:zeta_p}
\begin{align}
\tilde{c}_j(t)&\equiv e^{2\int_0^{t}K_{jj}(\tau)d\tau}c_j(t),\\
\tilde{K}_{j,\perp}(t)&\equiv e^{2\int_0^{t}K_{jj}(\tau)d\tau}K_{j,\perp}(t),
\end{align}
\end{subequations}
such that \cref{eq:c_eq_dif} becomes
\begin{subequations}\label{eq:c_tilde}
\begin{align}
\dot{z}_j(t)&=-2\left(\tilde{K}_{j,\perp}\tilde{c}_j^*+\tilde{K}^*_{j,\perp}\tilde{c}_j\right),\\
\dot{\tilde{c}}_j&=\tilde{K}_{j,\perp}z_j-\frac{\kappa}{2}\tilde{c}_j.
\end{align}
\end{subequations}
If we define $u_j\equiv \text{Re}(\tilde{K}_{j,\perp}\tilde{c}_j^*)$, then $\dot{z}_j=-4u_j$, so after some algebra with the equations in \cref{eq:c_tilde} and \cref{eq:zeta_p} we obtain 
\begin{equation}
\ddot{z}_j + \frac{\kappa}{2} \dot{z}_j + 4|\tilde{K}_{j,\perp}|^2 z_j = 0.
\end{equation}
The exact solution depends on $\nu_j\equiv \sqrt{(\kappa/2)^2-16|K_{j,\perp}|^2}$, which can lead to under- or overdamped behavior. However, in the strong-measurement regime
\begin{equation}
\kappa \gg |\tilde{K}_{j,\perp}| = ||B(t)||_2, 
\end{equation}
the slow mode approximately obeys
\begin{equation}\label{eq:Dustin}
\dot{z}_j \simeq -\frac{8}{\kappa} |\tilde{K}_{j,\perp}(t)|^2 z_j = \frac{8}{\kappa} ||B(t)||_2^2 z_j,
\end{equation}
so \cref{eq:Dustin} becomes
\begin{equation}
z_j(t)\simeq z_j(0)\exp\left[-\frac{8}{\kappa}\int_0^t ||B(\tau)||^2d\tau\right].
\end{equation}

Returning to the populations in terms of \cref{eq:zeta_p},
\begin{subequations}
\begin{align}
p_j(t) &= \frac{1}{2}(1+z_j(t)) , \\
q_j(t) &= \frac{1}{2}(1-z_j(t)) ,
\end{align}
\end{subequations}
we assume that at $t=0$ the state is in the codespace, so $p_j(0)=1$ while $q_j(0)=0$. The total probability of being in the codespace at the final time is
\begin{equation}
p_{\text{code}} \equiv p_0(t_f) + p_1(t_f),
\end{equation}
Therefore, we obtain the result in \cref{eq:leak_prob_cat}.

The tightest closed-form analytic bound we can write for the full infinite-dimensional oscillator is obtained by bounding the leakage block in \cref{eq:B_ortho}. The only non-negligible code-to-leakage source is the squeezing part, because $a^2$ stays inside the cat manifold, but the $a^{\dagger \, 2}$ part leaks out, up to exponentially small corrections:
\begin{equation}
||B(t)|| = \frac{1}{2} |\dot{r}(t)|\,||\mathbb{Q}_0 a^{\dagger 2} \mathbb{P}_0||_2 + \mathcal{O}(e^{-|\alpha|^2}).
\end{equation}
For the 4-cat code, the exact mean photon numbers in the logical cat states are
\begin{subequations}
\begin{align}
\bar{n}_0 &= |\alpha|^2 \frac{\sinh|\alpha|^2 - \sin|\alpha|^2 }{\cosh|\alpha|^2 + \cos |\alpha|^2},\\
\bar{n}_2 &= |\alpha|^2 \frac{\sinh|\alpha| + \sin|\alpha|^2 }{\cosh|\alpha|^2 - \cos |\alpha|^2}.
\end{align}
\end{subequations}
Hence, integrating the squared norm of $B$ we obtain
\begin{equation}\label{eq:tight}
\int_0^{t_f}\hspace{-0.3cm}||B(\tau)||^2d\tau=
\frac{j_{0,1}^2 \pi^2}{2|\alpha|^4 t_f}\max(4\bar{n}_0+2,4\bar{n}_2+2)+\mathcal{O}(e^{-2|\alpha|^2}).
\end{equation}
For large cats, both $\bar{n}_0,\bar{n}_2\rightarrow |\alpha|^2$, so an approximate analytical bound for the leakage probability becomes
\begin{equation}
1-p_{\mathrm{code}}(t_f) \lesssim 1 - \exp\left[-\frac{4 j_{0,1}^2\pi^2}{\kappa t_f|\alpha|^2} \left(1+\frac{1}{2|\alpha|^2}\right) \right], \nonumber
\end{equation}
which is a loose bound on \cref{eq:leak_prob_cat}, but is the tightest envelope that can be justified, as seen in the \cref{fig:leakage_cat}.

\section{Holonomic protocol for the GKP code}\label{app:WZ_GKP}

\subsection{Horizontal lift for the GKP code}

The dynamics of a state $\ket{\psi(t)}$ in the codespace is given in \cref{eq:psi_of_t}; the horizontal lift $L(t)$ is given in \cref{eq:L_of_t}, and the unitary $h(t)$ in \cref{eq:h_oft}. We wish to make $h(t_f)\approx \xi I$ a trivial phase, so we need to compute the term $V^{\dagger}V$ for the unitary rotation \cref{eq:man_of_the_year}:
\begin{equation}\label{eq:K_of_t_GKP}
V^\dagger(t)\dot V(t) = i\,\dot\kappa(t) \, f(Q/\sqrt{\pi} + g(t))
- i\sqrt{\pi}\,\dot g(t)\,P.
\end{equation}
Here, $f$ is the polynomial given in \cref{eq:f_gkp}. First, we show that any function $F(Q)$ is diagonal in the GKP basis defined by \cref{eq:0_1_GKP}, since the states are defined over the eigenstates of $Q$, $Q\ket{q} = q\ket{q}$:
\begin{equation}
\bra{q} F(Q) \ket{q} = F(q) \delta(q-q').
\end{equation}
So for $j$ and $k$ different integers,
\begin{align}
&\bra{2j\sqrt{\pi}} F(Q) \ket{(2k+1)\sqrt{\pi}} \nonumber\\
&= F(2j\sqrt{\pi}) \delta(2j\sqrt{\pi} - (2k+1)\sqrt{\pi}) = 0.
\end{align}
Hence, according to \cref{eq:0_1_GKP}, we obtain 
\begin{subequations}
\begin{align}
\bra{\mathbf{0}} F(Q) \ket{\mathbf{1}}& = 0 ,\\
\bra{\mathbf{1}} F(Q) \ket{\mathbf{0}}& = 0.
\end{align}
\end{subequations}
By contrast, the nonzero diagonal terms are
\begin{align}
\bra{\mathbf{0}} F(Q) \ket{\mathbf{0}} &= \sum_{j,k} \bra{2j\sqrt{\pi}} F(Q) \ket{2k\sqrt{\pi}} \nonumber\\
&= \sum_{j} F(2j\sqrt{\pi}),
\end{align}
and similarly,
\begin{equation}
\bra{\mathbf{1}} F(Q) \ket{\mathbf{1}} = \sum_{j,k} F((2j+1)\sqrt{\pi}).
\end{equation}
So by plugging the polynomial $F(Q)=f(Q/\sqrt{\pi}+g)$ into \cref{eq:f_gkp}, we obtain
\begin{subequations}
\begin{align}
\bra{\mathbf{0}}f(Q/\sqrt{\pi}+g)\ket{\mathbf{0}}&=f(g)\\
\bra{\mathbf{1}}f(Q/\sqrt{\pi}+g)\ket{\mathbf{1}}&=f(g+1),
\end{align}
\end{subequations}
where we ignored the modulo 2 values, $2j=0\mod 2$ and $2j+1=1\mod 2$, that are consequences of the GKP periodicity. Hence, we can write it as a logical operator
\begin{equation}
\mathbb{P}_{\text{GKP}} f(Q/\sqrt{\pi}+g) \mathbb{P}_{\text{GKP}} = a_I(t) I_L + a_Z(t) Z_L,
\end{equation}
where $\mathbb{P}_{\text{GKP}}$ denotes the codespace projector onto the GKP states into \cref{eq:0_1_GKP}, and 
\begin{subequations}
\begin{align}
a_I(t) &= \frac{f(g) + f(g+1)}{2} = \frac{g^3}{4} + \frac{g^2}{2} + \frac{g}{4} + \frac{1}{16}, \\
a_Z(t) &= \frac{f(g)-f(g+1)}{2} = -\frac{3g^2}{8} - \frac{g}{2} - \frac{1}{16}.
\end{align}
\end{subequations}

Now we go back to the Wilczek Zee term $V^{\dagger}\dot{V}$. First, the momentum variable $P$ acts as a leakage operator in the ideal GKP case:
\begin{equation}
\mathbb{P}_{\text{GKP}}P\mathbb{P}_{\text{GKP}} = 0.
\end{equation}
We take the easiest case in \cref{eq:kappa_t}, where $\dot{\kappa}$ is the constant $2\pi/t_f$.

We chose $g(t)$ in \cref{eq:g_t} such that $a_Z$ integrated from $0$ to $t_f$ is zero, so the $Z_L$ component vanishes:
\begin{equation}
\int_{0}^{t_f} a_Z(t) dt = -t_f \left(\frac{9A^2}{64} + \frac{A}{4} + \frac{1}{16}\right) = 0,
\end{equation}
where $A=-(8\pm 2\sqrt{7})/9$ (either choice of the sign works). For the coefficient of the $I_L$ term we get
\begin{align}
\int_{0}^{t_f} a_I(t) dt &= t_f \left(\frac{5 A^3}{64} + \frac{3A^2}{16} + \frac{A}{8} + \frac{1}{16}\right) \\
&= t_f \frac{437\mp 10\sqrt{7}}{11664}.
\end{align}
The unitary $h(t)$ given in \cref{eq:h_oft} becomes
\begin{equation}
h(t_f) = \exp\left(\frac{2\pi}{t_f} \int_{0}^{t_f} a_I(t) dt \right) L^{\dagger}(0) L(0),
\end{equation}
which acts as a trivial phase on the codespace. So we can conclude that at the final time the state is
\begin{equation}
\ket{\psi(t_f)} = V(t_f)L(0)h(t_f) \ket{\psi(0)} \propto T_{\text{GKP}} \ket{\psi(0)}.
\end{equation}

\subsection{Error-correcting conditions}\label{app:KL_gkp}

To evaluate the error-correcting conditions for the instantaneous codespace, we analyze
\begin{equation}
\mathbb{P}_{\mathrm{GKP}} V^{\dagger}(t) E_j^{\dagger} E_i V(t) \mathbb{P}_{\mathrm{GKP}},
\end{equation}
where the correctable error set has the form in \cref{eq:error_GKP}, the unitary is given by \cref{eq:man_of_the_year}, and they satisfy the conditions \cref{eq:Voronoi}. To simplify the notation we define
\begin{subequations}
\begin{align}
\Delta u &\equiv u_i-u_j,\\
\Delta v &\equiv v_i-v_j,
\end{align}
\end{subequations}
and from the position and momentum quadratures defined in \cref{eq:quadratures} we get
\begin{equation}
V^{\dagger}(t)QV(t)=Q+\sqrt{\pi}g(t).
\end{equation}
Since $Q$ and $P$ are conjugate quadratures satisfying $[Q,P]=i$, we get the commutator
\begin{equation}
[P,f(Q/\sqrt{\pi})]=-i\frac{1}{\sqrt{\pi}}f'(Q/\sqrt{\pi}).
\end{equation}
Using the property
\begin{equation}
e^{-A}Be^{A}=B+[B,A]+\frac{1}{2}[[B,A],A]+...,
\end{equation}
we obtain the factor
%\begin{subequations}
\begin{align}
V^{\dagger} P V &= e^{i\sqrt{\pi}gP} e^{-i\kappa f(Q/\sqrt{\pi})} P e^{i\kappa f(Q/\sqrt{\pi})} e^{-i\sqrt{\pi}P} \nonumber\\
&= e^{i\sqrt{\pi}P} \left[P + \frac{\kappa}{\sqrt{\pi}} f'(Q/\sqrt{\pi}) \right] e^{-i\sqrt{\pi}gP} \nonumber\\
&= P + \frac{\kappa}{\sqrt{\pi}} f'(Q/\sqrt{\pi}+g).
\end{align}
%\end{subequations}
So, substituting these two factors, we obtain
\begin{align}
&V^{\dagger}(t) E_{j}^{\dagger} E_i V(t)\\
&= e^{-iv_j \Delta u} (V^{\dagger} e^{-i\Delta u P} V) (V^{\dagger} e^{i\Delta v Q} V) \nonumber \\
&= e^{-iv_j\Delta u} e^{-i\Delta u \left(P+\frac{\kappa}{\sqrt{\pi}} f'(Q/\sqrt{\pi}+g) \right)} e^{i\Delta v(Q+\sqrt{\pi}g)}.\nonumber
\end{align}
If we sandwich this between ideal GKP projectors, we get
\begin{align}
\mathbb{P}_{\mathrm{GKP}} V^{\dagger}(t) E_{j}^{\dagger} E_i V(t) \mathbb{P}_{\mathrm{GKP}}
=& e^{-iv_j\Delta u} \mathbb{P}_{\mathrm{GKP}} \\
&\times e^{-i\Delta u \left(P + \frac{\kappa}{\sqrt{\pi}} f'(Q/\sqrt{\pi}+g)\right)} \nonumber\\
&\times e^{i\Delta v Q} \mathbb{P}_{\mathrm{GKP}} e^{i\Delta v\sqrt{\pi}g}. \nonumber
\end{align}
That is the exact sandwiched operator.
The useful reduction is:

If $\Delta u\notin\sqrt{\pi}\mathbb{Z}$, then the translation moves the GKP comb off the lattice, so that in the ideal code,
\begin{equation}
\mathbb{P}_{\mathrm{GKP}}V^{\dagger}(t)E_{j}^{\dagger}E_i V(t)\mathbb{P}_{\mathrm{GKP}}=0.
\end{equation}
If  $\Delta u=m\sqrt{\pi}$ and $\Delta v=n\sqrt{\pi}$ with $m$ and $n$ integers, then it reduces to a code-space operator; generally, a logical Pauli times a phase coming from the cubic dressing:
\begin{align}
&\mathbb{P}_{\mathrm{GKP}}V^{\dagger}(t)E_{j}^{\dagger}E_i V(t)\mathbb{P}_{\mathrm{GKP}}\nonumber\\&=e^{-iv_jm\sqrt{\pi}}M_{m,n}(t)X_L^mZ_L^n\mathbb{P}_{\mathrm{GKP}}. 
\end{align}
where $M_{m,n}(t)$ is the phase generated by
\begin{equation}
M_{m,n}(t)=e^{-i\Delta u \frac{\kappa(t)}{\sqrt{\pi}}f'(Q/\sqrt{\pi}+g(t))}.
\end{equation}
For the KL condition, the correctable set is the off-lattice, small-displacement region, so the sandwiched dressed error still vanishes for $i\neq j$ in the ideal code.

\subsection{Truncated GKP states}\label{app:truncated}

In position space, the ideal square-lattice GKP states described in \cref{eq:0_1_GKP} are the distributions
\begin{subequations}
\begin{align}
\psi_{\mathrm{ideal}}^{(k)}(q) &\equiv \braket{q|k_L} \propto \sum_{t\in \mathbb{Z}}\delta (q-q_t^{(k)})\\
q_t^{(k)} &\equiv (2t+k)\sqrt{\pi},
\end{align}
\end{subequations}
where $k\in\{0,1\}$ denotes $0_L$ or $1_L$, respectively.

A standard finite-energy regularization \cite{Menicucci_2014} with energy parameter $\epsilon$ is 
\begin{equation}
\ket{k_{\epsilon}}\propto e^{-\epsilon N}\ket{k_L},
\end{equation}
where $N$ is the number operator from \cref{eq:N_op}. We can represent $N$ in terms of the quadratures from \cref{eq:quadratures} as
\begin{equation}
N = \frac{1}{2}(Q^2+P^2-1).
\end{equation}
The operator $e^{-\epsilon N}$ acts like a Gaussian low-pass filter in phase space: it broadens each delta-function spike into a Gaussian wave packet and adds a Gaussian envelope across the comb. In position space, the truncated distribution is
\begin{equation}\label{eq:psi_epsilon_1}
\psi_{\epsilon}^{(k)} = \bra{Q} e^{-\epsilon N} \ket{k_L} \propto \sum_{t\in \mathbb{Z}} \bra{Q} e^{-\epsilon N} \ket{Q_t^{(k)}} .
\end{equation}

The matrix elements of $e^{-\epsilon N}$ in the position basis are the kernels in the Mehler representation:
\begin{align}
\bra{Q} e^{-\epsilon N} \ket{Q_t} =& \frac{1}{\sqrt{\pi(1-e^{-2\epsilon})}} \\
&\times \exp\bigg[-\frac{(Q^2+(Q_t)^2)\cosh \epsilon-2QQ_t}{2\sinh \epsilon}\bigg], \nonumber
\end{align}
where we have omitted the superscript $(k)$ in $q_t$. Later we are going to insert the ideal GKP points
\begin{equation}
    Q_t\equiv Q_t^{(k)}=(2t+k)\sqrt{\pi}.
\end{equation}
So each delta-function peak is mapped to a Gaussian function in $Q$, and after completing the square \cref{eq:psi_epsilon_1} becomes
%\begin{subequations}
\begin{align} \label{eq:psi_epsilon_2}
\psi_{\epsilon}^{(k)} &= \mathcal{N}_{\epsilon} \sum_{t\in \mathbb{Z}} \exp\bigg[-\frac{\pi}{2} \tanh \epsilon (2t+k)^2 \bigg] \psi_t^{(k)}(Q), \nonumber\\ 
\psi_t^{(k)}(Q) &\equiv \exp\bigg[-\frac{\left(Q-\frac{\sqrt{\pi}(2t+k)}{\cosh \epsilon}\right)^2}{2\tanh \epsilon} \bigg],
\end{align}
%\end{subequations}
which is a sum of Gaussian wave packets centered at the lattice points, multiplied by a Gaussian envelope as written in \cref{eq:truncated_GKP}.

We can recognize each term in \cref{eq:psi_epsilon_2} as a displaced, squeezed vacuum state. A squeezed vacuum state in $Q$-space has a Gaussian width 
\begin{equation}
(\Delta Q)^2 = \frac{e^{-2r}}{2},
\end{equation}
and the displacement shifts the center by
\begin{equation}\label{eq:bar_Q}
\bar{Q} = \sqrt{2}\alpha.
\end{equation}
Then the displaced, squeezed vacuum state takes the form
\begin{equation}
\bra{Q} D(\alpha) S(r) \ket{0} \propto \exp \bigg[-\frac{(Q-\sqrt{2}\alpha)^2}{2e^{-2r}}\bigg].
\end{equation}
Finally, comparing this last equation with \cref{eq:psi_epsilon_2}, we get the expressions given in \cref{eq:truncated_GKP}.

There is one more thing to define before using simulations with the Strawberry Fields Python library, which is the parameter \text{ampl\_cutoff}$=1\times 10 ^{-12}$, which makes the lattice cutoff
\begin{equation}
z_{\rm max} = \left\lceil \sqrt{-\frac{\log(\mathrm{ampl\_cutoff})}{4\pi \tanh \epsilon}}\right\rceil.
\end{equation}

To get a bound on the phase error for the $T_{\mathrm{GKP}}$ gate in \cref{eq:delta_theta}, first we get the matrix elements from \cref{eq:M_t_gkp}. Substituting the finite-energy states, these are
\begin{equation}\label{eq:M_kk_state}
M_{kk} = \mathcal{N}_{\epsilon}^2 \sum_{t,t'} c_t^{(k)} c_{t'}^{(k)} K_{t,t'} c_t^{(k)} c_{t'}^{(k)} K_{tt'}^{(k)},
\end{equation}
where 
\begin{subequations}\begin{align}
K_{tt'}^{(k)} &= \bra{0} S^{\dagger} D^{\dagger} (\alpha_t^{(k)}) T_{\mathrm{GKP}} D(\alpha_{t'}^{(k)}) S\ket{0} \\
&= \int_{-\infty}^{\infty} dQ (\psi_{t}^{(k)}(Q))^{*} e^{i2\pi f(Q/\sqrt{\pi})} \psi_{t'}^{(k)}(Q).
\end{align}
\end{subequations}
Here, we used the fact that $T_{\mathrm{GKP}}$ in \cref{eq:T_GKP} is diagonal in the $Q-$basis, and \cref{eq:psi_epsilon_2}.
The overlap between distinct wave packets is
\begin{equation}
\braket{\psi_t|\psi_{t'}} \sim \exp\left[-\frac{(Q-\bar{Q})^2}{2\tanh \epsilon} \right],
\end{equation}
hence, the term in \cref{eq:M_kk_state} can be written as the sum just of the diagonal terms $t=t'$:
\begin{equation}\label{eq:trunc_Airy}
M_{kk} = \mathcal{N}_{\epsilon}^{2} \sum_{t} |c_t^{(k)}|^2 K_{tt}^{(k)} + \mathcal{O}\left(e^{-\pi/\tanh\epsilon}\right).
\end{equation}
To solve the integral $K_{tt}^{(k)}$, we change variables:
\begin{equation}
\xi_{Q}\equiv Q-\bar{Q},
\end{equation}
such that the diagonal term of the kernel is
\begin{equation}
K_{tt}^{(k)} \hspace{-0.1cm} = \hspace{-0.1cm} \int_{-\infty}^{\infty} \hspace{-0.2cm} d\xi_Q \exp\left(-\frac{\xi^2_Q}{2\tanh \epsilon}\right) \exp\bigg[i2\pi f\left(\frac{\bar{Q} + \xi_Q}{\sqrt{\pi}}\right)\bigg].
\end{equation}
Using \cref{eq:bar_Q} and \cref{eq:alpha_t_k}, we get
\begin{equation}
x_t \equiv \frac{\bar{Q}}{\sqrt{\pi}} = \frac{2t+k}{\cosh \epsilon}.
\end{equation}
Expanding the phase as a Taylor series, we get
\begin{align}
2\pi f\!\!\left(\!x_t+\!\frac{\xi_Q}{\sqrt{\pi}}\!\right)\! =\!2\pi f(x_t) + 2\sqrt{\pi}f'(x_t)\xi_Q \nonumber\\ &+ f''(x_t) \xi_Q^2 + \frac{\xi_Q^3}{2\sqrt{\pi}},
\end{align}
we can reduce the cubic-polynomial kernel integral in terms of an Airy function,
\begin{equation}
K_{tt}^{(k)} = e^{i2\pi f(x_t)} \int_{-\infty}^{\infty} d\xi_Q \exp\left(a_t\xi_Q^3+b_t\xi_Q^2+c_t\xi_Q\right),
\end{equation}
with
\begin{equation}
a_t = \frac{i}{2\sqrt{\pi}},\quad b_t = if''(x_t) - \frac{1}{2\tanh\epsilon}, \quad c_t = i2\sqrt{\pi} f'(x_t).
\end{equation}
To eliminate the quadratic term, we define
\begin{equation}
\xi_Q \equiv y_Q - \frac{b_t}{3a_t}.
\end{equation}
The kernel becomes
\begin{align}
K_{tt}^{(k)} &= e^{i2\pi f(x_t)+\Gamma_t} \int_C dy e^{a_ty^3+\zeta_ty} \nonumber\\
&= e^{i2\pi f(x_t) + \Gamma_t} \frac{2\pi}{(3a_t)^{1/3}} \mathrm{Ai}\left(-\frac{\zeta_t}{(3a_t)^{1/3}}\right),
\end{align}
where
\begin{equation}
\zeta_t \equiv c_t - \frac{b_t^2}{2a_t},\qquad \Gamma_t = \frac{2b_t^3}{27a_t^2} - \frac{b_tc_t}{3a_t}.
\end{equation}

We can find an  analytical approximation for the smooth phase error $\delta\theta$ using \cref{eq:delta_theta}, with the diagonal matrix elements $M_{kk}$ from \cref{eq:trunc_Airy}. And finally, we can derive the \textit{leakage probability} due to measurement,
\begin{equation}
p_{\rm leak} = \mathrm{Tr}(\mathbb{Q}_{\epsilon}\rho),
\end{equation}
using the rotating-frame Lindblad master equation with the truncated projector $\mathbb{P}_{\epsilon}$ in \cref{eq:Proj_trunc_GKP}:
%\begin{subequations}
\begin{align}
\frac{dp_{\rm leak}}{ds} &= -\mathrm{Tr}\{\mathbb{Q}_{\epsilon} [V^{\dagger}(s)\dot{V}(s),\rho] \} \nonumber\\
&= -\mathrm{Tr}(\mathbb{Q}_{\epsilon} V^{\dagger} \dot{V} \mathbb{P}_\epsilon \rho) + \mathrm{Tr}(\mathbb{Q}_{\epsilon} \rho  \mathbb{P}_\epsilon V^{\dagger} \dot{V}) \nonumber\\
&= -2\mathrm{Tr}(\mathbb{Q}_{\epsilon} V^{\dagger} \dot{V} \mathbb{P}_\epsilon \rho),
\end{align}
%\end{subequations}
where we used the fact that the dissipator does not directly change the leakage probability:
\begin{equation}
\mathrm{Tr}(\mathbb{Q}_\epsilon \mathcal{D}[\mathbb{P}_\epsilon]) = 0.
\end{equation}
Since $\mathbb{Q}_\epsilon^2=\mathbb{Q}_\epsilon$,$\mathbb{P}_\epsilon^2=\mathbb{P}_\epsilon$, and defining $\rho_{PQ}=\mathbb{P}_\epsilon \rho \mathbb{Q}_\epsilon$,
\begin{align}
\frac{dp_{\rm leak}}{ds} &= -2\mathrm{Tr}(\mathbb{Q}_{\epsilon} V^{\dagger} \dot{V} \mathbb{P}_\epsilon \rho) \nonumber\\
&= -2\mathrm{Tr}(\mathbb{Q}_{\epsilon} V^{\dagger} \dot{V} \mathbb{P}_\epsilon \rho_{PQ}).
\end{align}

In the strong-measurement regime, the coherence block closely follows the Hamiltonian coupling. Solving the damped equation to leading order gives the estimate
\begin{equation}
||\rho_{PQ}||_2 \leq \frac{2}{\kappa} ||\mathbb{Q}_\epsilon  V^{\dagger} \dot{V} \mathbb{P}_\epsilon||_2.
\end{equation}
Then in the leakage rate:
\begin{align}
\frac{dp_{\rm leak}}{ds} &\leq 2 ||\mathbb{Q}_\epsilon V^{\dagger} \dot{V} \mathbb{P}_\epsilon||_2 \times ||\rho_{PQ}||_2 \nonumber\\
&\leq \frac{4}{\kappa} ||\mathbb{Q}_\epsilon V^{\dagger} \dot{V} \mathbb{P}_\epsilon||^2_2.
\end{align}
Integrating from $s=0$ to $1$ we obtain the inequality in \cref{eq:bound_leak_GKP}.

\end{document}